\documentclass[11pt,a4paper]{article}
\usepackage{amssymb}
\usepackage{graphicx}
\usepackage{amsmath}
\usepackage{makeidx}
\usepackage{indentfirst}
\usepackage[T1]{fontenc}
\usepackage[utf8]{inputenc}
\usepackage{authblk}

\newcounter{resultnum}[section]
\newcounter{conclusionnum}[section]
\newcounter{conditionnum}[section]
\newcounter{conjecturenum}[section]
\newcounter{examplenum}[section]
\newcounter{exercisenum}[section]
\newcounter{lemmanum}[section]
\newcounter{notationnum}[section]
\newcounter{theoremnum}[section]
\newcounter{definitionnum}[section]
\newcounter{corollarynum}[section]
\newcounter{remarknum}[section]
\newcounter{propositionnum}[section]
\newcounter{acknowledgementnum}[section]
\newcounter{algorithmnum}[section]
\newcounter{axiomnum}[section]
\newcounter{casenum}[section]
\newcounter{claimnum}[section]
\newcounter{summarynum}[section]
\newcounter{problemnum}[section]
\begin{document}

\title{Exact Solutions in Modified Massive Gravity and Off--Diagonal
Wormhole Deformations}
\date{April 1, 2013 and September 10, 2013}

\author[1]{Sergiu I. Vacaru\thanks{sergiu.vacaru@cern.ch;\ sergiu.vacaru@uaic.ro} }

\affil[1]{\small Theory Division, CERN, CH-1211, Geneva 23, Switzerland \footnote{associated visiting researcher
\newline
The paper was accepted to EJPC. On March 3, 2014, arXiv moderators considered author’s appeal and permitted to resubmit this work to qr-qc.  Originally, the manuscript was submitted on April 1, 2013 as arXiv submit/686664’, on 29 pages. Moderators requested to substitute it as a new version of a "short" 4 pages preprint arXiv: 13033515v1 which was accepted for publication by other editor.
};\ and
\newline
 Rector's Office, Alexandru Ioan Cuza University,\
 Alexandru Lapu\c sneanu street, \newline  nr. 14, UAIC - Corpus R, office 323;\
 Ia\c si,\ Romania, 700057
}
\maketitle

\begin{abstract}
There are explored off--diagonal deformations of "prime" metrics in Einstein
gravity (for instance, for wormhole configurations) into "target" exact
solutions in f(R,T)--modified and massive/bi--metric gravity theories. The
new classes of solutions may posses, or not, Killing symmetries and can be
characterized by effective induced masses, anisotrop\-ic polarized
interactions and cosmological constants. For nonholonomic deformations with
(conformal) ellipsoid/ toroid and/or solitonic symmetries and, in
particular, for small eccentricity rotoid configurations, we can generate
wormholes like objects matching external black ellipsoid -- de Sitter
geometries. We conclude that there are nonholonomic tansforms and/or
non--trivial limits to exact solutions in general relativity when modified/
massive gravity effects are modelled by off--diagonal and/or nonholonomic
parametric interactions.

\vskip0.1cm


\vskip3pt

\end{abstract}

\affil[1]{\small Rector's Office, Alexandru Ioan Cuza University, \newline
Alexandru Lapu\c sneanu street, nr. 14, UAIC -- Corpus R, office
323;\newline  Ia\c si,\ Romania, 700057 }\renewcommand\Authands{ and }


\section{Introduction}

The bulk of physically important exact solutions in gravity theories (for
instance, defining black holes and wormholes) are described by metrics with
two Killing symmetries, see summaries of results in monographs \cite%
{kramer,griff}. For such solutions, there are certain "canonical" frames of
reference, when the coefficients of fundamental geometric/physical objects
depend generically on one or two (from maximum four, in four dimensional,
4--d, theories) spacetime coordinates. This class of metrics can be
diagonalized by  coordinate transforms, or contain off--diagonal terms
generated by rotations. To construct generic off--diagonal solutions
parameterized by metrics with six independent coefficients depending
generically on three and/or, in general, on all spacetime coordinates is a
very difficult technical and geometric task and the physical meaning of such
generalized/modified, or Einstein, spacetimes is less clear.

In our works, see \cite{vadm1,vadm2} and references therein, we elaborated a
geometric method which allows us to deform nonholonomically any "prime"
diagonal metric into various classes of "target" off--diagonal solutions
with one Killing and/or non-Killing symmetries. For deformations on a small
parameter, the new classes of target solutions may preserve certain
important physical properties of a prime metric (for instance, of a black
hole/ring one, or for a wormhole) but posses also new characteristics
related to anisotropic polarizations of constants, nonlinear off--diagonal
interactions with new symmetries etc.

Wormhole configurations with spacetime handles (shortcuts), nontrivial
topology and exotic matter \cite{morris} attract attention for theoretical
probes of foundations of gravity theories and as possible objects of nature
(for reviews, see \cite{visser,lobo,boehm} and references therein). Such
solutions are determined in reverse direction when some tunneling metrics of
prescribed (for instance, spherical and/or conformal) symmetry are
considered and then try to find some corresponding exotic matter sources. A
number of interesting and/or peculiar solutions was found when time like
curves and respective causality violations are allowed, for strass--energy
tensors with possible violation of the null energy conditions. The wormhole
subjects where revived some times in connection to black hole solutions,
coupling with gauge interactions, singularities, generalized/modified
gravity theories etc.

We studied locally anisotropic wormhole and/or flux tubes in five
dimensional (5--d) gravity \cite{vw1,vw2,vw3}. Such objects can be
determined by extra dimensional or warped/ trapped configurations and/or \
possible ellipsoidal, toroidal, bipolar, solitonic etc gravitational
polarizations of vacuum and/or gravitational constants. The corresponding
spacetime geometries are described by generic off--diagonal metrics\footnote{%
which can not be diagonalized by coordinate transforms} with coefficients
depending on three or four coordinates and various types of (pseudo)
Riemannian or non--Riemannian connections.

In the present paper, we address the problem of constructing deformations of
prime wormhole metrics in general relativity, GR, resulting in generic
off--diagonal solutions in modified gravity, MG, and theories with
nonholonomically induced torsion, effective masses and bi--metric and
bi--connection structures. We shall work with two equivalent connections
(the Levi--Civita and an auxiliary one) defined by the same metric structure
and apply and extend the anholonomic frame deformation method (AFDM, see
details in \cite{vadm1,vadm2}, and references therein) of constructing exact
solutions in gravity theories.

The idea of the AFDM is to find certain classes of nonholonomic
(equivalently, anholonomic/ non-integrable) frames with conventional
2+2+..., or 3+2+..., splitting of dimensions on (pseudo) Riemannian
spacetime when the (in general, modified) Einstein equations decouple for a
correspondingly defined "auxiliary" connection. This results into systems of
nonlinear partial differential equations (PDE) which can be integrated in
very general forms. The corresponding solutions are with generic
off--diagonal metrics and generalized connections. They may depend on all
spacetime coordinates via generating and integration functions. The
formalism is different from that with more "simple" diagonal ansatz when the
Einstein equations are transformed into certain systems of nonlinear
ordinary differential equations (ODE). For instance, for the second order
ODE, we get only integration constants which are related to certain physical
ones like the gravitational constant, a point particle mass and/or electric
charge etc following certain asymptotic/boundary conditions.

We argue that it is possible to impose such constraints on a nonholonomic
frame structure, via corresponding classes of generating/integration
functions, when the "auxiliary" torsion vanishes and we can "extract"
solutions for the Einstein gravity theory and various modifications. To
provide a physical interpretation of certain off--diagonal exact solutions
with one--Killing or non--Killing symmetries is usually a very difficult
task. In general, it is not clear if any physical meaning/importance can be
found for a derived new class of generalized solutions. Nevertheless, it is
possible to elaborate realistic physical models with nonhlonomically
constrained nonlinear off--diagonal gravitational and matter field
interactions if we consider deformations on a small parameter (for instance,
small eccentricities for ellipsoid/rotoid configurations). This allows us to
construct new classes of off--diagonal solutions determining parametric
deformations of wormhole and black hole physical objects resulting into new
observable physical effects and more complex spacetime configurations.

The article is organized as follows: We formulate a geometric approach to
modified massive gravity theories in section \ref{s2}. A proof that
corresponding gravitational field equations can be decoupled and integrated
in general forms with respect to certain classes of nonholonomic frames of
references is provided in section \ref{s3}. The method of off--diagonal
deformations of wormhole -- de Sitter configurations is outlined in section %
\ref{s4}. There are considered small parametric deformations resulting in
physically interesting solutions. In section \ref{s5}, four classes of
"locally anisotropic" deformations of original wormhole metrics are
constructed. We deduce spacetime metrics for rotoid deformations of
wormholes, consider solitonic waves on such wormholes and (if possible?)
black ellipsoids and explore a model with a torus ringing the throat of a
wormhole. In a more general context, massive gravity and $f$--modifications
to configurations with nonholonomically induced (by metric coefficients)
torsions are considered. Section \ref{s6} is devoted to concluding remarks.

\section{Nonholonomic Deformations in Modified Massive Gravity}

\label{s2} We outline certain geometric methods on nonholonomic 2+2
spacetime splitting provided in detail in Refs. \cite{vadm1,vadm2}.

\subsection{Geometric preliminaries}

We shall refer to gravity theories formulated on a four dimensional, 4--d, \
generalized pseudo--Riemannian manifold $\mathbf{V}$ endowed with metric
structure $\mathbf{g}$ and a metric compatible linear connection $\mathbf{D}%
, $ $\mathbf{Dg}=0$. There will be considered distortion relations of type
\begin{equation}
\mathbf{D}=\nabla +\mathbf{Z},  \label{distr}
\end{equation}%
when both the "auxiliary", $\mathbf{D}$, and Levi--Civita, $\nabla =\{\Gamma
_{\ \beta \gamma }^{\alpha }\},$ connections and the distortion tensor, $%
\mathbf{Z}=\{\mathbf{Z}_{\ \beta \gamma }^{\alpha }\},$ are completely
defined by the coefficients $\mathbf{g}=\{g_{\alpha \beta }(u^{\gamma })\}.$
To construct a natural splitting (\ref{distr}) following a well--defined
geometric principle we can introduce a conventional horizontal (h) and
vertical (v) splitting of the tangent space $T\mathbf{V,}$ when a
non--integrable (equivalently, nonholonomic, or anholonomic) distribution
\begin{equation}
\mathbf{N}:T\mathbf{V}=h\mathbf{V}\oplus v\mathbf{V}  \label{whit}
\end{equation}%
is determined locally via a set of coefficients $\mathbf{N}%
=\{N_{i}^{a}(x,y)\}$; a 2+2 splitting can be parameterized by local
coordinates $u=(x,y)$, $u^{\mu }=(x^{i},y^{a}),$ where indices run values $%
i,j,...=1,2$ and $a,b,...=3,4.$\footnote{
The coefficients $\Gamma _{\ \beta \gamma }^{\alpha },Z_{\ \beta \gamma
}^{\alpha }$ and $g_{\alpha \beta }$ are computed with respect to certain
(co) frames of reference, $e_{\alpha }=e_{\ \alpha }^{\alpha ^{\prime
}}(u)\partial _{\alpha ^{\prime }}$ and $e^{\beta }=e_{\ \beta ^{\prime
}}^{\beta }(u)du^{\beta ^{\prime }},$ for $\partial _{\alpha ^{\prime
}}:=\partial /\partial u^{\alpha ^{\prime }}$. The Einstein rule on
summation on \textquotedblright up-low\textquotedblright\ cross indices will
be applied if the contrary is not stated. On convenience, "primed",
"underlined" etc indices will be used. The local pseudo--Euclidean signature
is fixed in the form $(+++-).$ We shall write boldface letters in order to
emphasize that a nonlinear connection, N--connection, structure (\ref{whit})
is fixed on spacetime manifold $\mathbf{V.}$}

A h--v--splitting (\ref{whit}) results in a structure of N--adapted\ local
bases, $\mathbf{e}_{\nu }=(\mathbf{e}_{i},e_{a})$, and cobases, $\mathbf{e}%
^{\mu}=(e^{i},\mathbf{e}^{a}),$ when
\begin{eqnarray}
\mathbf{e}_{i} &=&\partial /\partial x^{i}-\ N_{i}^{a}(u)\partial /\partial
y^{a},\ e_{a}=\partial _{a}=\partial /\partial y^{a},  \label{nader} \\
\mbox{ and  }e^{i} &=&dx^{i},\ \mathbf{e}^{a}=dy^{a}+\ N_{i}^{a}(u)dx^{i}.
\label{nadif}
\end{eqnarray}%
For such frames, there are satisfied the nonholonomy relations
\begin{equation}
\lbrack \mathbf{e}_{\alpha },\mathbf{e}_{\beta }]=\mathbf{e}_{\alpha }%
\mathbf{e}_{\beta }-\mathbf{e}_{\beta }\mathbf{e}_{\alpha }=W_{\alpha \beta
}^{\gamma }\mathbf{e}_{\gamma },  \label{nonholr}
\end{equation}%
with nontrivial anholonomy coefficients
\begin{equation}
W_{ia}^{b}=\partial _{a}N_{i}^{b},W_{ji}^{a}=\Omega _{ij}^{a}=\mathbf{e}%
_{j}\left( N_{i}^{a}\right) -\mathbf{e}_{i}(N_{j}^{a}).  \label{anhcoef}
\end{equation}

We can distinguish the coefficients of geometric objects on $\mathbf{V}$
with respect to N--adapted (co) frames (\ref{nader}) and (\ref{nadif}) and
call them, in brief, d--objects. For instance, a vector $Y(u)\in T\mathbf{V}
$ can be parameterized as a d--vector, $\mathbf{Y}=$ $\mathbf{Y}^{\alpha }%
\mathbf{e}_{\alpha }=\mathbf{Y}^{i}\mathbf{e}_{i}+\mathbf{Y}^{a}e_{a},$ or $%
\mathbf{Y}=(hY,vY),$ with $hY=\{\mathbf{Y}^{i}\}$ and $vY=\{\mathbf{Y}%
^{a}\}. $

Any metric structure on $\mathbf{V}$ can be written (up to general
frame/coordinate transforms) in two equivalent forms: with respect to a dual
local coordinate basis,
\begin{equation*}
\mathbf{g}=\underline{g}_{\alpha \beta }du^{\alpha }\otimes du^{\beta },
\end{equation*}%
where
\begin{equation}
\underline{g}_{\alpha \beta }=\left[
\begin{array}{cc}
g_{ij}+N_{i}^{a}N_{j}^{b}g_{ab} & N_{j}^{e}g_{ae} \\
N_{i}^{e}g_{be} & g_{ab}%
\end{array}%
\right]  \label{ansatz}
\end{equation}%
or as a d--metric,
\begin{equation}
\mathbf{g}=g_{\alpha }(u)\mathbf{e}^{\alpha }\otimes \mathbf{e}^{\beta
}=g_{i}(x)dx^{i}\otimes dx^{i}+g_{a}(x,y)\mathbf{e}^{a}\otimes \mathbf{e}%
^{a}.  \label{dm1}
\end{equation}

On a nonholonomic manifold $(\mathbf{V,N}),$ we can consider a subclass of
linear connections called \textit{distinguished connections, d--connections,}
$\mathbf{D}=(hD,vD),$ preserving under parallelism the N--connection
splitting (\ref{whit}). Any $\mathbf{D}$ defines an operator of covariant
derivative, $\mathbf{D}_{\mathbf{X}}\mathbf{Y}$, for a d--vector field $%
\mathbf{Y}$ in the direction of a d--vector $\mathbf{X}.$ With respect to
N--adapted frames (\ref{nader}) and (\ref{nadif}), the value $\mathbf{D}_{%
\mathbf{X}}\mathbf{Y}$ can be computed as in GR but with the coefficients of
the Levi--Civita connection substituted by $\mathbf{D}=\{\mathbf{\Gamma }_{\
\alpha \beta }^{\gamma }=(L_{jk}^{i},L_{bk}^{a},C_{jc}^{i},C_{bc}^{a})\}$.
The respective coefficients are computed for the h--v--components of $%
\mathbf{D}_{\mathbf{e}_{\alpha }}\mathbf{e}_{\beta }:=$ $\mathbf{D}_{\alpha }%
\mathbf{e}_{\beta }$ using $\mathbf{X}=\mathbf{e}_{\alpha }$ and $\mathbf{Y}=%
\mathbf{e}_{\beta }.$

A d--connection is characterized by three fundamental geometric objects: the
d--torsion, $\mathcal{T},$ the nonmetricity, $\mathcal{Q},$ and the
d--curvature, $\mathcal{R},$ all defined by standard formulas
\begin{eqnarray*}
\mathcal{T}(\mathbf{X,Y}) &:=&\mathbf{D}_{\mathbf{X}}\mathbf{Y}-\mathbf{D}_{%
\mathbf{Y}}\mathbf{X}-[\mathbf{X,Y}],\mathcal{Q}(\mathbf{X}):=\mathbf{D}_{%
\mathbf{X}}\mathbf{g,} \\
\mathcal{R}(\mathbf{X,Y}) &:=&\mathbf{D}_{\mathbf{X}}\mathbf{D}_{\mathbf{Y}}-%
\mathbf{D}_{\mathbf{Y}}\mathbf{D}_{\mathbf{X}}-\mathbf{D}_{\mathbf{[X,Y]}}.
\end{eqnarray*}%
We can compute the corresponding N--adapted coefficients,
\begin{eqnarray*}
\mathcal{T} &=&\{\mathbf{T}_{\ \alpha \beta }^{\gamma }=\left( T_{\
jk}^{i},T_{\ ja}^{i},T_{\ ji}^{a},T_{\ bi}^{a},T_{\ bc}^{a}\right) \},%
\mathcal{Q}=\mathbf{\{Q}_{\ \alpha \beta }^{\gamma }\}, \\
\mathcal{R} &\mathbf{=}&\mathbf{\{R}_{\ \beta \gamma \delta }^{\alpha }%
\mathbf{=}\left( R_{\ hjk}^{i}\mathbf{,}R_{\ bjk}^{a}\mathbf{,}R_{\ hja}^{i}%
\mathbf{,}R_{\ bja}^{c}, R_{\ hba}^{i},R_{\ bea}^{c}\right)\},
\end{eqnarray*}%
of these geometric objects by introducing $\mathbf{X}=\mathbf{e}_{\alpha }$
and $\mathbf{Y}=\mathbf{e}_{\beta },$ and $\mathbf{D}=\{\mathbf{\Gamma }_{\
\alpha \beta }^{\gamma }\}$ into above formulas, see details in \cite%
{vadm1,vadm2}.

It should be noted that the Levi--Civita connection $\nabla $ (in brief, LC,%
\footnote{%
it is uniquely defined by the metric structure $\mathbf{g}$ if there are
imposed two conditions:\ $\mathcal{T}=0 $ and $\mathcal{Q}=0$, if $\mathbf{%
D\rightarrow \nabla}$}) is not a d--connection because it does not preserve
under general frame/coordinate transforms the N--connection splitting (\ref%
{whit}). Nevertheless,\textit{\ there is a canonical d--connection} $%
\widehat{\mathbf{D}}$ also uniquely determined by any geometric data $(%
\mathbf{g},\mathbf{N}) $ following two similar but a bit "relaxed"
conditions: 1) it is metric compatible, $\widehat{\mathbf{D}}\mathbf{g=0,}$
and 2) with zero h--torsion, $h\widehat{\mathcal{T}}=\{\widehat{T}_{\
jk}^{i}\}=0,$ and zero v--torsion, $v\widehat{\mathcal{T}}=\{\widehat{T}_{\
bc}^{a}\}=0.$ This allows us to construct a canonical distortion relation of
type (\ref{distr}) with respective splitting of N--adapted coefficients $%
\widehat{\mathbf{\Gamma }}_{\ \alpha \beta }^{\gamma }=\Gamma _{\ \alpha
\beta }^{\gamma }+\widehat{\mathbf{Z}}_{\ \alpha \beta }^{\gamma }$. We can
work equivalently with two metric compatible connections $\widehat{\mathbf{D}%
}$ and $\mathbf{\nabla }$ because both such geometric objects are completely
defined by the same metric structure $\mathbf{g.}$\footnote{%
The N--adapted coefficients of $\widehat{\mathbf{D}}=\{$ $\widehat{\mathbf{%
\Gamma }}_{\ \alpha \beta }^{\gamma }=(\widehat{L}_{jk}^{i},\widehat{L}%
_{bk}^{a},\widehat{C}_{jc}^{i},\widehat{C}_{bc}^{a})\}$ and $\widehat{%
\mathbf{Z}}_{\ \alpha \beta }^{\gamma }$ depending only on $g_{\alpha \beta
} $ and $N_{i}^{a}$ can be computed following formulas\
\begin{eqnarray}
\widehat{L}_{jk}^{i} &=&\frac{1}{2}g^{ir}\left( \mathbf{e}_{k}g_{jr}+\mathbf{%
e}_{j}g_{kr}-\mathbf{e}_{r}g_{jk}\right), \widehat{C}_{bc}^{a}=\frac{1}{2}%
g^{ad}\left( e_{c}g_{bd}+e_{b}g_{cd}-e_{d}g_{bc}\right)  \notag \\
\widehat{C}_{jc}^{i} &=&\frac{1}{2}g^{ik}e_{c}g_{jk},\ \widehat{L}%
_{bk}^{a}=e_{b}(N_{k}^{a})+\frac{1}{2}g^{ac}\left( \mathbf{e}%
_{k}g_{bc}-g_{dc}\ e_{b}N_{k}^{d}-g_{db}\ e_{c}N_{k}^{d}\right),  \notag
\end{eqnarray}%
see proofs, for instance, in \cite{vadm1,vadm2}.} For the canonical
d--connection, there are nontrivial d--torsions coefficients, {\small
\begin{equation}
\widehat{T}_{\ jk}^{i}=\widehat{L}_{jk}^{i}-\widehat{L}_{kj}^{i},\widehat{T}%
_{\ ja}^{i}=\widehat{C}_{jb}^{i},\widehat{T}_{\ ji}^{a}=-\Omega _{\ ji}^{a},%
\widehat{T}_{aj}^{c}=\widehat{L}_{aj}^{c}-e_{a}(N_{j}^{c}),\widehat{T}_{\
bc}^{a}=\ \widehat{C}_{bc}^{a}-\ \widehat{C}_{cb}^{a}.  \label{dtors}
\end{equation}%
} The geometric meaning of such a nonholonomically induced torsion is
different from that, for instance, in Riemann--Cartan geometry because in
our approach $\widehat{\mathcal{T}}$ is completely defined by the metric
structure.

A (pseudo) Riemannian geometry can be formulated alternatively in "geometric
variables" $(\mathbf{g,N,}\widehat{\mathbf{D}})$ computing in standard form,
respectively, the Riemann, $\widehat{\mathcal{R}}\mathbf{=}\mathbf{\{}%
\widehat{\mathbf{R}}_{\ \beta \gamma \delta }^{\alpha }\},$ and the Ricci, $%
\widehat{\mathcal{R}}ic=\{\widehat{\mathbf{R}}_{\ \beta \gamma }\},$
d--tensors. For instance, the nonsymmetric d--tensor $\widehat{\mathbf{R}}%
_{\alpha \beta }:=\widehat{\mathbf{R}}_{\ \alpha \beta \gamma }^{\gamma }$
of $\widehat{\mathbf{D}}$ is characterized by four $h$--$v$ N--adapted
coefficients{\small
\begin{equation}
\widehat{\mathbf{R}}_{\alpha \beta }=\{\widehat{R}_{ij}:=\widehat{R}_{\
ijk}^{k},\ \widehat{R}_{ia}:=-\widehat{R}_{\ ika}^{k},\ \widehat{R}_{ai}:=%
\widehat{R}_{\ aib}^{b},\ \widehat{R}_{ab}:=\widehat{R}_{\ abc}^{c}\},
\label{driccic}
\end{equation}%
}which allows us to compute an "alternative" scalar curvature
\begin{equation}
\ \widehat{R}:=\mathbf{g}^{\alpha \beta }\widehat{\mathbf{R}}_{\alpha \beta
}=g^{ij}\widehat{R}_{ij}+g^{ab}\widehat{R}_{ab}.  \label{sdcurv}
\end{equation}%
We can also introduce the Einstein d--tensor of $\widehat{\mathbf{D}},$
\begin{equation}
\widehat{\mathbf{E}}_{\alpha \beta }\doteqdot \widehat{\mathbf{R}}_{\alpha
\beta }-\frac{1}{2}\mathbf{g}_{\alpha \beta }\ \widehat{R}.  \label{enstdt}
\end{equation}

The values $\widehat{\mathcal{R}},\widehat{\mathcal{R}}ic$ and $\ \widehat{R}
$\ for the canonical d--connection $\widehat{\mathbf{D}}$ are different from
the similar ones, $\mathcal{R},\mathcal{R}ic$ and $R,$ computed for the
LC--connection $\nabla $. Nevertheless, both classes of such fundamental
geometric objects are related via unique distorting relations derived from (%
\ref{distr}) for a N--connection splitting (\ref{whit}). To work with $%
\widehat{\mathbf{D}}$ is convenient for various purposes in generalized
gravity theories with nontrivial torsion. The most surprising property of
the Ricci d--tensor $\widehat{\mathcal{R}}ic=\{\widehat{\mathbf{R}}_{\ \beta
\gamma }\}$ is that the corresponding modified Einstein equations of type $%
\widehat{\mathbf{R}}_{\ \beta \gamma }=\Upsilon _{\ \beta \gamma }$ decouple
in very general forms with respect to certain classes of N--adapted frames.
This property holds for generic off--diagonal ansatz of type (\ref{ansatz})
(in principle, depending on all coordinates) and for certain formally
diagonalized and N--adapted sources $\Upsilon _{\ \beta \gamma }.$ This
allows us to generate various classes of exact solutions in commutative and
noncommutative gravity theories with 4--d and higher dimensions spacetimes,
see details and examples in Refs. \cite{vadm1,vadm2,vw1,vw2,vw3}. Such a
geometric method of constructing exact solutions in gravity is
conventionally called the anholonomic frame deformation method (AFDM).

The AFDM can be used for constructing off--diagonal exact solutions in
general relativity (GR) and other theories involving the LC--connection $%
\mathbf{\nabla .}$ In such cases, $\widehat{\mathbf{D}}=\{\widehat{\mathbf{%
\Gamma }}_{\ \alpha \beta }^{\gamma }\}$ can be considered as an "auxiliary"
connection which together with certain convenient sets of N--coefficients, $%
N_{i}^{a}$, are introduced with the aim to decouple certain systems of
nonlinear partial differential equations (PDE) and solve them in very
general forms. Such solutions are determined by corresponding classes of
generating and integration functions and, in principle, on an infinite
number of integration/symmetry parameters. On corresponding integral
varieties of solutions, we can impose additional nonholonomic constraints
when the torsion (\ref{dtors}) vanishes and $\widehat{\mathbf{D}}\rightarrow
\nabla .$ Such constraints result in first order PDE equations which can be
of type
\begin{equation}
\widehat{L}_{aj}^{c}=e_{a}(N_{j}^{c}),\widehat{C}_{jb}^{i}=0,\Omega _{\
ji}^{a}=0,  \label{lccond}
\end{equation}%
These equations can be solved also in very general forms and allows us to
extract LC--configurations. We note that if we work from the very beginning
with $\mathbf{\nabla}$, we can not decouple for general off--diagonal
metrics, for instance, the Einstein equations. This is a consequence of
generic nonlinearity of gravitational filed equations. The priority of $%
\widehat{\mathbf{D}}$ is that we can "relax" a bit the zero torsion
conditions, decouple the corresponding nonlinear PDEs for certain convenient
systems of reference determined by "flexible" $N_{i}^{a}$ and find general
classes of solutions. At the end (after a class of generalized metrics and
connections was defined), we can constrain nonholonomically/parametrically
the nonlinear system and find torsionless configurations.

The main goal of this work is to show that the AFDM allows us to generate
exact solutions with nonholonomic deformations of wormhole objects in
modified and/or massive gravity.

\subsection{Nonholonomic massive f(R,T) gravity}

We study modified gravity theories derived for the action%
\begin{equation}
S=\frac{1}{16\pi }\int \delta u^{4}\sqrt{|\mathbf{g}_{\alpha \beta }|}[f(%
\widehat{R},T)-\frac{\mu _{g}^{2}}{4}\mathcal{U}(\mathbf{g}_{\mu \nu },%
\mathbf{K}_{\alpha \beta })+\ ^{m}L].  \label{act}
\end{equation}%
Such theories generalize the so--called modified $f(R,T)$ gravity, see
reviews and original results in \cite{odints1,odints2,odints3}, and the
ghost--free massive gravity (by de Rham, Gabadadze and Tolley, dRGT) \cite%
{drg1,drg2,drg3}. This evades from certain problems of the bi--metric theory
by Hassan and Rosen, \cite{hr1,hr2} and connects us to various recent
research in black hole physics and modern cosmology \cite{nieu,koyam,muko}.
In this paper, we shall use the units when $\hbar =c=1$ and the Planck mass $%
M_{Pl}$ is defined via $M_{Pl}^{2}=1/8\pi G$ with 4--d Newton constant $G.$
We write $\delta u^{4}$ instead of $d^{4}u$ because there are used
N--elongated differentials (\ref{nader}) and consider the constant $\mu _{g}$
as the mass parameter for gravity. The geometric and physical meaning of the
values contained in this formula will be explained below.

There are at least three most important motivations to consider in this work
such generalized models of gravity: 1) Using nonholonomic deformations
described in previous section, we can transform certain classes of solutions
in modified gravity into certain equivalent ones for massive gravity. 2) Via
off--diagonal gravitational interactions in Einstein gravity, it is possible
to mimic various classes of physical effects in modified, massive,
bi--metric and bi--connection gravity. 3) The AFDM seems to be an effective
geometric tool for constructing exact solutions in such "sophisticate"
gravity theories.

In action (\ref{act}), the Lagrangian density $\ ^{m}L$ is used for
computing the stress--energy tensor of matter via variation in N--adapted
form, using operators (\ref{nader}) and (\ref{nadif}), on inverse metric
d--tensor (\ref{dm1}),  $\mathbf{T}_{\alpha \beta }=-\frac{2}{\sqrt{|\mathbf{%
g}_{\mu \nu }|}}\frac{\delta (\sqrt{|\mathbf{g}_{\mu \nu }|}\ ^{m}L)}{\delta
\mathbf{g}^{\alpha \beta }}$, when the trace is computed $T:=\mathbf{g}%
^{\alpha \beta }\mathbf{T}_{\alpha \beta }.$ The functional $f(\widehat{R},T)
$ modifies the standard Einstein--Hilbert Lagrangian (with $R$ for the
Levi--Civita connection $\nabla )$ to that for the modified $f$--gravity but
with dependence on $\ ^{s}\widehat{R}$ (\ref{sdcurv}) and $T.$ In a large
class of generalized cosmological models, we can assume that the
stress--energy tensor of the matter is given by
\begin{equation}
\mathbf{T}_{\alpha \beta }=(\rho +p)\mathbf{v}_{\alpha }\mathbf{v}_{\beta }-p%
\mathbf{g}_{\alpha \beta },  \label{emt}
\end{equation}%
for the approximation of perfect fluid matter with the energy density $\rho $
and the pressure $p$; the four--velocity $\mathbf{v}_{\alpha }$ being
subjected to the conditions $\mathbf{v}_{\alpha }\mathbf{v}^{\alpha }=1$ and
$\mathbf{v}^{\alpha }\widehat{\mathbf{D}}_{\beta }\mathbf{v}_{\alpha }=0,$
for $\ ^{m}L=-p$ in a corresponding local N--adapted frame. For simplicity,
we can parametrize
\begin{equation}
f(\widehat{R},T)=\ ^{1}f(\widehat{R})+\ ^{2}f(T)  \label{functs}
\end{equation}%
and denote by $\ ^{1}F(\widehat{R}):=\partial \ ^{1}f(\widehat{R})/\partial
\widehat{R}$ and $\ ^{2}F(T):=\partial \ ^{2}f(T)/\partial T.$

In addition to the usual $f$--gravity term (in particular, to the
Einstein--Hilbert one) in (\ref{act}), it is considered a mass term with
"gravitational mass" $\mu _{g}$ and potential {\small
\begin{eqnarray}
\mathcal{U}/4 &=&-12+6[\sqrt{\mathcal{S}}]\mathcal{+[S}]\mathcal{-[}\sqrt{%
\mathcal{S}}]^{2}+  \label{potent} \\
&&\alpha _{3}\{18[\sqrt{\mathcal{S}}]-6[\sqrt{\mathcal{S}}]^{2}+[\sqrt{%
\mathcal{S}}]^{3}+2\mathcal{[S}^{3/2}]-3\mathcal{[S}]([\sqrt{\mathcal{S}}%
]-2)-24\}+  \notag \\
&&\alpha _{4}\{[\sqrt{\mathcal{S}}](24-12\mathcal{[}\sqrt{\mathcal{S}}]-%
\mathcal{[}\sqrt{\mathcal{S}}]^{3})-12[\sqrt{\mathcal{S}}]\mathcal{[S}]+2%
\mathcal{[}\sqrt{\mathcal{S}}]^{2}(3\mathcal{[S}]+2\mathcal{[}\sqrt{\mathcal{%
S}}])+  \notag \\
&&3\mathcal{[S}](4-\mathcal{[S}])-8\mathcal{[S}^{3/2}](\sqrt{\mathcal{S}}%
-1)+6\mathcal{[S}^{2}]-24\},  \notag
\end{eqnarray}%
} where the trace of a matrix $\mathcal{S}=(S_{\mu \nu })$ is dentoted by $%
\mathcal{[S}]:=S_{\ \nu }^{\nu };$ the square root of such a matrix, $\sqrt{%
\mathcal{S}}=(\sqrt{\mathcal{S}}_{\ \mu }^{\nu }),$ is understood to be a
matrix for which $\sqrt{\mathcal{S}}_{\ \alpha }^{\nu }\sqrt{\mathcal{S}}_{\
\mu }^{\alpha }=S_{\ \mu }^{\nu }$ and $\alpha _{3}$ and $\alpha _{4}$ are
free parameters. This nonlinearly extended Fierz--Pauli type potential was
shown to result in a theory of massive gravity which is free from
ghost--like degrees of freedom and takes a special form of total derivative
in absence of dynamics (see \cite{drg2,drg3} and additional arguments in
\cite{gratia}). The potential generating matrix $\mathcal{S}$ is constructed
in a special form to result in a d--tensor $\mathbf{K}_{\ \mu }^{\nu
}=\delta _{\ \mu }^{\nu }-\sqrt{\mathcal{S}}_{\ \mu }^{\nu }$ characterizing
metric fluctuations away from a fiducial (flat) 4--d spacetime. The
coefficients
\begin{equation}
\mathbf{S}_{\ \mu }^{\nu }=\mathbf{g}^{\nu \alpha }\eta _{\overline{\nu }%
\overline{\mu }}\mathbf{e}_{\alpha }s^{\overline{\nu }}\mathbf{e}_{\mu }s^{%
\overline{\mu }},  \label{stuk}
\end{equation}
with the Minkowski metric $\eta _{\overline{\nu }\overline{\mu }%
}=diag(1,1,1,-1),$ are generated by introducing four scalar St\"{u}kelberg
fileds $s^{\overline{\nu }},$ which is necessary for restoring the
diffeomorphism invariance. Using N--adapted values $\mathbf{g}^{\nu \alpha }$
and $\mathbf{e}_{\alpha }$ we can always transform a tensor $S_{\mu \nu }$
into d--tensor $\mathbf{S}_{\mu \nu }$ characterizing nonholonomically
constrained fluctuations. This is possible for the values $\mathbf{K}_{\ \mu
}^{\nu },\mathbf{S}_{\ \mu }^{\nu },\sqrt{\mathcal{S}}_{\ \mu }^{\nu }$ etc
even $s^{\overline{\nu }}$ transforms as scalar fields under coordinate and
frame transforms.

Varying the action (\ref{act}) in N--adapted from for the coefficients of
d--metric $\mathbf{g}_{\nu \alpha }$ (\ref{dm1}), we obtain certain
effective Einstein equations, see (\ref{enstdt}), for the modified massive
gravity,%
\begin{equation}
\widehat{\mathbf{E}}_{\alpha \beta }=\mathbf{\Upsilon }_{\beta \delta },
\label{efcdeq}
\end{equation}%
with source
\begin{equation}
\mathbf{\Upsilon }_{\beta \delta }=\ ^{ef}\eta \ G\ \mathbf{T}_{\beta \delta
}+\ ^{ef}\mathbf{T}_{\beta \delta }+\mu _{g}^{2}\ ^{K}\mathbf{T}_{\beta
\delta }.  \label{effectsource}
\end{equation}%
The first component in such a source is determined by usual matter fields
with energy momentum $\mathbf{T}_{\beta \delta }$ tensor but with effective
polarization of the gravitational constant $\ ^{ef}\eta =[1+\ ^{2}F/8\pi ]/\
^{1}F $. The $f$--modification of the energy--momentum tensor also results
in the section term as an additional effective source {\small
\begin{equation}
\ ^{ef}\mathbf{T}_{\beta \delta }=[\frac{1}{2}(\ ^{1}f-\ ^{1}F\ \widehat{R}%
+2p\ ^{2}F+\ ^{2}f)\mathbf{g}_{\beta \delta }-(\mathbf{g}_{\beta \delta }\
\widehat{\mathbf{D}}_{\alpha }\widehat{\mathbf{D}}^{\alpha }-\widehat{%
\mathbf{D}}_{\beta }\widehat{\mathbf{D}}_{\delta })\ ^{1}F]/\ ^{1}F
\label{efm}
\end{equation}%
} and "mass gravity" contribution (the third term) is computed as a
dimensionless effective stress--energy tensor {\small
\begin{eqnarray*}
&&\ ^{K}\mathbf{T}_{\alpha \beta }:=\frac{1}{4\sqrt{|\mathbf{g}_{\mu \nu }|}}%
\frac{\delta (\sqrt{|\mathbf{g}_{\mu \nu }|}\ \mathcal{U})}{\delta \mathbf{g}%
^{\alpha \beta }} \\
&=&-\frac{1}{12}\{\ \mathcal{U}\mathbf{g}_{\alpha \beta }/4-2\mathbf{S}%
_{\alpha \beta }+2([\sqrt{\mathcal{S}}]-3)\sqrt{\mathcal{S}}_{\alpha \beta }+
\\
&&\alpha _{3}[3(-6+4\mathcal{[}\sqrt{\mathcal{S}}]+\mathcal{[}\sqrt{\mathcal{%
S}}]^{2}-\mathcal{[S}])\sqrt{\mathcal{S}}_{\alpha \beta }+6(\mathcal{[}\sqrt{%
\mathcal{S}}]-2)\mathbf{S}_{\alpha \beta }-\mathcal{S}_{\alpha \beta
}^{3/2}]- \\
&&\alpha _{4}[24\left( \mathcal{S}_{\alpha \beta }^{2}-([\sqrt{\mathcal{S}}%
]-1)\mathcal{S}_{\alpha \beta }^{3/2}\right) ]+12(2-2[\sqrt{\mathcal{S}}]-%
\mathcal{[S}]+[\sqrt{\mathcal{S}}]^{2})\mathbf{S}_{\alpha \beta }+ \\
&&(24-24[\sqrt{\mathcal{S}}]+12[\sqrt{\mathcal{S}}]^{2}-[\sqrt{\mathcal{S}}%
]^{3}-12[\mathcal{S}]+12[\mathcal{S}][\sqrt{\mathcal{S}}]-8\mathcal{[S}%
^{3/2}])\sqrt{\mathcal{S}}_{\alpha \beta }\}.
\end{eqnarray*}%
}In "hidden" form, $\ ^{K}\mathbf{T}_{\alpha \beta }$ encode a bi--metric
configuration with the second (fiducial) d--metric \textbf{\ }$\mathbf{f}%
_{\alpha \mu }=\eta _{\overline{\nu }\overline{\mu }}\mathbf{e}_{\alpha }s^{%
\overline{\nu }}\mathbf{e}_{\mu }s^{\overline{\mu }}$ determined by the St%
\"{u}kelberg fields $s^{\overline{\nu }}.$ The potential $\mathcal{U}$ (\ref%
{potent}) defines interactions between $\mathbf{g}_{\mu \nu }$ and $\mathbf{f%
}_{\mu \nu }$ via $\sqrt{\mathcal{S}}_{\ \mu }^{\nu }=\sqrt{\mathbf{g}^{\nu
\mu }\mathbf{f}_{\alpha \nu }}$ and $\mathcal{S}_{\ \mu }^{\nu }:=\mathbf{g}%
^{\nu \mu }\mathbf{f}_{\alpha \nu }.$ For simplicity, we shall study in this
paper bi--metric gravity models with $\ ^{K}\mathbf{T}_{\alpha \beta }=\
\lambda (x^{k})\ \mathbf{g}_{\alpha \beta },$ which can be generated by such
$s^{\overline{\nu }}$ when $\mathbf{g}_{\mu \nu }=\iota ^{2}(x^{k})\mathbf{f}%
_{\mu \nu }$ up to a nontrivial conformal factor $\varpi ^{2}.$ Using (\ref%
{stuk}), we can compute $\mathcal{S}_{\ \mu }^{\nu }:=\iota ^{-2}\delta _{\
\mu }^{\nu }$ which allows to express the effective polarized anisotropic
constant$\ $encoding the contributions of $s^{\overline{\nu }}$ as a
functional $\lambda \lbrack \iota ^{2}(x^{k})].$ In general, the solutions
of (\ref{efcdeq}) depend on the type of symmetries of interactions we
prescribe for $\mathbf{f}_{\alpha \mu }$ which, in our model, are N--adapted
and subjected to additional nonholonomic constraints.

The gravitational field equations (\ref{efcdeq}) are similar to the Einstein
ones in GR but for a different metric compatible linear connection, $%
\widehat{\mathbf{D}},$ and with nonlinear "gravitationally polarized"
coupling in effective source $\mathbf{\Upsilon }_{\beta \delta }$ (\ref%
{effectsource}). Such nonlinear systems of PDE can be integrated in general
forms for any N--adapted parameterizations {\small
\begin{equation}
\mathbf{\Upsilon }_{~\delta }^{\beta }=diag[\mathbf{\Upsilon }_{\alpha }:%
\mathbf{\Upsilon }_{~1}^{1}=\mathbf{\Upsilon }_{~2}^{2}=\Upsilon
(x^{k},y^{3});\mathbf{\Upsilon }_{~3}^{3}=\mathbf{\Upsilon }%
_{~4}^{4}=~^{v}\Upsilon (x^{k})],  \label{source}
\end{equation}%
}in particular, if
\begin{equation}
\Upsilon =~^{v}\Upsilon =\Lambda =const,  \label{source1}
\end{equation}%
for an effective cosmological constant $\Lambda ,$ see details in \cite%
{vadm1,vadm2}. A solution of equations (\ref{efcdeq}) for a source (\ref%
{source}) can be modelled effectively by certain classes of solutions
generated by a N--adapted constant coefficients (\ref{source1}) if the
generating and integration functions are redefined to mimic certain classes
of solutions. This is equivalent to a procedure of fixing a value for the
auxiliary scalar curvature $\widehat{R}$ (\ref{sdcurv}) by frame/coordinate
transforms of $N_{i}^{a}$ and related N--adapted bases which holds true not
for arbitrary 2+2 splitting but for certain classes of nonholonomic frames
resulting in decoupling of the generalized Einstein equations and necessary
parameterizations for sources. Here we note that $\widehat{\mathbf{D}}%
_{\delta }\ ^{1}F_{\mid \Upsilon =\Lambda }=0$ in (\ref{efm}) if we
prescribe a functional dependence $\ \widehat{R}=const.$ For rather general
distributions of matter fields and effective matter, we can prescribe such
values for (\ref{source1}) with $\mathbf{T}_{\beta \delta }=\check{T}(x^{k})%
\mathbf{g}_{\beta \delta }$ and $\ ^{s}\widehat{R}=\widehat{\Lambda }$ in (%
\ref{source}),%
\begin{eqnarray}
\Upsilon &=&\Lambda =\ ^{ef}\eta \ G\ \check{T}(x^{k})+  \label{source1a} \\
&&\frac{1}{2}(\ ^{1}f(\widehat{\Lambda })-\widehat{\Lambda }\ ^{1}F(\widehat{%
\Lambda })\ +2p\ ^{2}F(\check{T})+\ ^{2}f(\check{T}))+\mu _{g}^{2}\ \lambda
(x^{k}),  \notag \\
\ ^{ef}\eta &=&[1+\ ^{2}F(\check{T})/8\pi ]/\ ^{1}F(\widehat{\Lambda }).
\notag
\end{eqnarray}%
In general, any term may depend on coordinates $x^{i}$ but via
re--definition of generating functions they can be transformed into certain
effective constants, see bellow the footnote \ref{fnoteredif}. Prescribing
values $\widehat{\Lambda },\check{T},\ \lambda ,p$ and functionals $\ ^{1}f$
and $\ ^{2}f,$ we describe a nonholonomic matter and effective matter fields
dynamics with respect to N--adapted frames.

Finally, we note that the effective source $\mathbf{\Upsilon }_{~\delta
}^{\beta }=\Lambda \mathbf{\delta }_{~\delta }^{\beta }$ (via nonholonomic
constraints and the canonical d--connection $\widehat{\mathbf{D}}$) encode
all information on modifications of the GR theory to certain classes of $f$%
--modified and/or massive gravity theories. Imposing additional constraints
when $\widehat{\mathbf{D}}_{\mathcal{T}=0}\rightarrow \nabla ,$ i.e. solving
the equations (\ref{lccond}), we extract LC--configurations for above
mentioned gravitational models.

\section{Decoupling \& Integrability of MG Field Eqs}

\label{s3} In this section, we formulate and analyze possible conditions on
the nonholonomic frame structure and matter fields and effective matter
distributions when the gravitational field equations for f--modified
bi--metric field equations decouple and can be integrated in very general
forms. We show that such generic off--diagonal solutions depend on various
classes of generating and integration functions and parameters. Such
modified spacetimes describe nonholonomic deformations of a prime (fiducial
and/or well defined metric in GR, for simplicity, taken in a diagonal form
with two Killing symmetries) into certain "target" configurations in
modified gravity theories.

There are analyzed three classes of target solutions: 1) nonvacuum
off--diagonal deformations to Levi--Civita configurations with effective
cosmological constants encoding contributions from massive and $f$--modified
gravity; 2) possible generalizations to nontrivial nonholonomically induced
torsion configurations; and 3) nonholonomic deformations on a small
parameter.

\subsection{Decoupling with respect to N--adapted frames}

The local coordinates on a 4--d manifold $\mathbf{V}$ are parameterized in
the form $u^{\mu }=(x^{i},y^{a})=(x^{1}=r,x^{2}=\theta ,y^{3}=\varphi
,y^{4}=t)$ (or, in brief, $u=(x,y)$), where indices run values $i,j,...=1,2$
and $a,b,...=3,4$ and $\ t$ is a timelike coordinate. In brief, the partial
derivatives $\partial _{\alpha }=\partial /\partial u^{\alpha }$ will be
labeled in the forms $s^{\bullet }=\partial s/\partial x^{1},s^{\prime
}=\partial s/\partial x^{2},s^{\ast }=\partial s/\partial y^{3},s^{\diamond
}=\partial s/\partial y^{4}.$

We shall study nonholonomic deformations of a prime metric\footnote{%
we consider that such a metric is with two Killing vector symmetries and
that in certain systems of coordinates it can be diagonalized}
\begin{eqnarray*}
\mathbf{\mathring{g}} &=&\mathring{g}_{\alpha }(u)\mathbf{\mathring{e}}%
^{\alpha }\otimes \mathbf{\mathring{e}}^{\beta }=\mathring{g}%
_{i}(x)dx^{i}\otimes dx^{i}+\mathring{g}_{a}(x,y)\mathbf{\mathring{e}}%
^{a}\otimes \mathbf{\mathring{e}}^{a}, \\
\mbox{ for } &&\mathbf{\mathring{e}}^{\alpha }=(dx^{i},\mathbf{e}^{a}=dy^{a}+%
\mathring{N}_{i}^{a}(u)dx^{i}), \\
&&\mathbf{\mathring{e}}_{\alpha }=(\mathbf{\mathring{e}}_{i}=\partial
/\partial y^{a}-\mathring{N}_{i}^{b}(u)\partial /\partial y^{b},\ {e}%
_{a}=\partial /\partial y^{a}),
\end{eqnarray*}%
into a target off--diagonal one
\begin{eqnarray}
\mathbf{g} &=&g_{\alpha }(u)\mathbf{e}^{\alpha }\otimes \mathbf{e}^{\beta
}=g_{i}(x)dx^{i}\otimes dx^{i}+g_{a}(x,y)\mathbf{e}^{a}\otimes \mathbf{e}^{a}
\label{dm} \\
&=&\eta _{i}(x^{k})\mathring{g}_{i}dx^{i}\otimes dx^{i}+\eta
_{a}(x^{k},y^{b})\mathring{h}_{a}\mathbf{e}^{a}\otimes \mathbf{e}^{a},
\notag
\end{eqnarray}%
where $\mathbf{e}^{a}$ are taken as in (\ref{nadif}).\ Our goal is to
generate $\mathbf{g}$ as an exact solution in a (modified) gravity theory
even $\mathbf{\mathring{g}}$ is not obligatory constrained to the condition
to be a solution of any gravitational field equations. For certain
bi--metric models, the prime metric $\mathbf{\mathring{g}}$ can be
considered as a fiducial one which via nonholonomic nonlinear gravitational
interactions results in a solution in modified/ massive gravity. In next
sections, we shall take $\mathbf{\mathring{g}}$ as a wormhole solution in GR
and study possible off--diagonal deformations induced in generalized gravity
theories. We shall study the conditions when modified gravity effects can be
explained alternatively by certain effective nonlinear interactions in GR.

The nontrivial components of the Einstein equations (\ref{efcdeq}) with
source (\ref{source}) parameterized with respect to N--adapted bases (\ref%
{nader}) and (\ref{nadif}) for a metric ansatz (\ref{dm}) with data (\ref%
{data2}) for $\omega =1$ are{\small
\begin{eqnarray}
-\widehat{R}_{1}^{1} &=&-\widehat{R}_{2}^{2}=\frac{1}{2g_{1}g_{2}}%
[g_{2}^{\bullet \bullet }-\frac{g_{1}^{\bullet }g_{2}^{\bullet }}{2g_{1}}-%
\frac{\left( g_{2}^{\bullet }\right) ^{2}}{2g_{2}}+g_{1}^{\prime \prime }-%
\frac{g_{1}^{\prime }g_{2}^{\prime }}{2g_{2}}-\frac{(g_{1}^{\prime })^{2}}{%
2g_{1}}]=\ ^{v}\Upsilon ,  \label{eq1b} \\
-\widehat{R}_{3}^{3} &=&-\widehat{R}_{4}^{4}=\frac{1}{2h_{3}h_{4}}%
[h_{4}^{\ast \ast }-\frac{\left( h_{4}^{\ast }\right) ^{2}}{2h_{4}}-\frac{%
h_{3}^{\ast }h_{4}^{\ast }}{2h_{3}}]=\Upsilon ,  \label{eq2b} \\
\widehat{R}_{3k} &=&\frac{w_{k}}{2h_{4}}[h_{4}^{\ast \ast }-\frac{\left(
h_{4}^{\ast }\right) ^{2}}{2h_{4}}-\frac{h_{3}^{\ast }h_{4}^{\ast }}{2h_{3}}%
]+\frac{h_{4}^{\ast }}{4h_{4}}(\frac{\partial _{k}h_{3}}{h_{3}}+\frac{%
\partial _{k}h_{4}}{h_{4}})-\frac{\partial _{k}h_{4}^{\ast }}{2h_{4}}=0,
\label{eq3b} \\
\widehat{R}_{4k} &=&\frac{h_{4}}{2h_{3}}n_{k}^{\ast \ast }+(\frac{h_{4}}{%
h_{3}}h_{3}^{\ast }-\frac{3}{2}h_{4}^{\ast })\frac{n_{k}^{\ast }}{2h_{3}}=0,
\label{eq4b}
\end{eqnarray}%
}when the torsionless (Levi--Civita, LC) conditions (\ref{lccond}) transform
into
\begin{eqnarray}
w_{i}^{\ast } &=&(\partial _{i}-w_{i}\partial _{3})\ln \sqrt{|h_{3}|}%
,(\partial _{i}-w_{i}\partial _{3})\ln \sqrt{|h_{4}|}=0,  \label{lccondb} \\
\partial _{k}w_{i} &=&\partial _{i}w_{k},n_{i}^{\ast }=0,\partial
_{i}n_{k}=\partial _{k}n_{i}.  \notag
\end{eqnarray}%
Proofs of such formulas (but for other types of sources in GR and
commutative and noncommutative Finsler like generalizations) are contained
in Refs. \cite{vadm1,vadm2}. The above system of nonlinear PDE posses an
important decoupling property which allows us to integrate step by step such
equations.

\subsection{Generating off--diagonal solutions}

We can integrate the Einstein equations (\ref{efcdeq}) for a source (\ref%
{source}) if the N--adapted coefficients of a metric (\ref{dm}) are
parameterized in the form
\begin{equation}
g_{i}=e^{\psi {(x^{k})}},g_{a}=\omega (x^{k},y^{b})h_{a}(x^{k},y^{3}),\
N_{i}^{3}=w_{i}(x^{k},y^{3}),N_{i}^{4}=n_{i}(x^{k}),  \label{data2}
\end{equation}%
considering that using frame/coordinate transforms we can satisfy the
conditions $h_{a}^{\ast }\neq 0,\Upsilon _{2,4}\neq 0.$ In a more general
context, it is possible to consider any class of metrics which via frame and
coordinate transforms can be related to such an ansatz. For
parameterizations (\ref{data2}), the system (\ref{eq1b})--(\ref{eq4b})
transforms correspondingly into
\begin{eqnarray}
\psi ^{\bullet \bullet }+\psi ^{\prime \prime } &=&2~^{v}\Upsilon
\label{eq1} \\
\phi ^{\ast }h_{4}^{\ast } &=&2h_{3}h_{4}\Upsilon  \label{eq2} \\
\beta w_{i}-\alpha _{i} &=&0,  \label{eq3} \\
n_{i}^{\ast \ast }+\gamma n_{i}^{\ast } &=&0,  \label{eq4} \\
\partial _{i}\omega -(\partial _{i}\phi /\phi ^{\ast })\omega ^{\ast
}-n_{i}\omega ^{\diamond } &=&0,  \label{confeq}
\end{eqnarray}%
for
\begin{equation}
\alpha _{i}=h_{4}^{\ast }\partial _{i}\phi ,\beta =h_{4}^{\ast }\ \phi
^{\ast },\gamma =\left( \ln |h_{4}|^{3/2}/|h_{3}|\right) ^{\ast },
\label{abc}
\end{equation}%
where
\begin{equation}
{\phi =\ln |h_{4}^{\ast }/\sqrt{|h_{3}h_{4}|}|}  \label{genf}
\end{equation}%
is considered as a generating function. The equation (\ref{confeq}) is
necessary if we introduce a nontrivial conformal (in the vertical
"subspace") factor depending on all four coordinates. It will be convenient
to work also with the value $\Phi :=e^{{\phi }}.$

The above systems of nonlinear PDE can be integrated step by step in very
general forms following such a procedure:

\begin{enumerate}
\item The (\ref{eq1}) is just a 2--d Laplace equation which allows us to
find $\psi $ for any given source $~^{v}\Upsilon .$

\item For $h_{a}:=\epsilon _{a}z_{a}^{2}(x^{k},y^{3}),$ when $\epsilon
_{a}=\pm 1$ depending on signature (we do not consider summation on
repeating indices in this formula), the equations (\ref{eq2}) and (\ref{genf}%
) are written correspondingly in the form%
\begin{equation}
\phi ^{\ast }z_{4}^{\ast }=\epsilon _{3}z_{4}(z_{3})^{2}\Upsilon \mbox{ and }%
e^{{\phi }}z_{3}=2\epsilon _{4}z_{4}^{\ast }.  \label{aux3}
\end{equation}%
Multiplying both equations for nonzero $z_{4}^{\ast },\phi ^{\ast },z_{a}$
and introducing the result instead of the first equation, this system
transforms into
\begin{equation}
\Phi ^{\ast }=2\epsilon _{3}\epsilon _{4}z_{3}z_{4}\Upsilon \mbox{ and }\Phi
z_{3}=2\epsilon _{4}z_{4}^{\ast }.  \label{aux3a}
\end{equation}%
Taking $z_{3}$ from the second equation and introducing in the first one, we
obtain $\lbrack (z_{4})^{2}]^{\ast }=\frac{\epsilon _{3}[\Phi ^{2}]^{\ast }}{%
4\Upsilon }$. This allows us to integrate on $y^{3}$ and write%
\begin{equation}
h_{4}=\epsilon _{4}(z_{4})^{2}=\ ^{0}h_{4}(x^{k})+\frac{\epsilon
_{3}\epsilon _{4}}{4}\int dy^{3}\frac{[\Phi ^{2}]^{\ast }}{\Upsilon },
\label{h4a}
\end{equation}%
for an integration function $\ ^{0}h_{4}(x^{k}).$\footnote{\label{fnoteredif}%
We can always re--define a generating function $\Phi
(x^{k},y^{3})\rightarrow \tilde{\Phi}(x^{k},y^{3})$ and a source $\Upsilon
(x^{k},y^{3})\rightarrow \Lambda ,$ reconsidering (\ref{aux3a}), in a form
when $[\Phi ^{2}]^{\ast }/4\Upsilon =$ $[\tilde{\Phi}^{2}]^{\ast }/4\Lambda
, $ which allows us to perform a formal integration in (\ref{h4a}) and get $%
h_{4}=\ ^{0}h_{4}(x^{k})+\epsilon _{3}\epsilon _{4}[\tilde{\Phi}^{2}]^{\ast
}/4\Lambda .$} Using the first equation in (\ref{aux3}), we find%
\begin{equation}
h_{3}=\epsilon _{3}(z_{3})^{2}=\frac{\phi ^{\ast }}{\Upsilon }\frac{%
z_{4}^{\ast }z_{4}}{z_{4}z_{4}}=\frac{1}{2\Upsilon }(\ln |\Phi |)^{\ast
}(\ln |h_{4}|)^{\ast }.  \label{h3a}
\end{equation}%
For $\Upsilon =\Lambda ,$ we can redefine the coordinates and $\Phi ,$
introduce $\epsilon _{3}\epsilon _{4}$ in $\Lambda $ and consider solutions
of type%
\begin{equation}
h_{3}[\Phi ]=(\Phi ^{\ast })^{2}/\Lambda \Phi ^{2}\mbox{ and }h_{4}[\Phi
]=\Phi ^{2}/4\Lambda ,  \label{h4bl}
\end{equation}

\item We have to solve algebraic equations for $w_{i}$ by introducing the
coefficients (\ref{abc}) in (\ref{eq3}) for the generating function $\phi ,$
or using any equivalent variables $\phi,\Phi ,$ and/or $\tilde{\Phi}$,
\begin{equation}
w_{i}=\partial _{i}\phi /\phi ^{\ast }=\partial _{i}\Phi /\Phi ^{\ast }.
\label{w1bl}
\end{equation}

\item The solution of the equation (\ref{eq4}) can be obtained by
integrating two times on $y^{3},$%
\begin{equation}
n_{k}=\ _{1}n_{k}+\ _{2}n_{k}\int dy^{3}\ h_{3}/(\sqrt{|h_{4}|})^{3},
\label{n1b}
\end{equation}%
where $\ _{1}n_{k}(x^{i}),\ _{2}n_{k}(x^{i})$ are integration functions.

\item The LC--conditions (\ref{lccondb}) consist a set of nonholonomic
constraints which can not be solved in explicit form for arbitrary data $%
(\Phi ,\Upsilon )$ and all types of integration functions $\ _{1}n_{k}$ and $%
\ _{2}n_{k}.$ Nevertheless, we can find explicit solutions if we consider
that via frame and coordinate transforms we can chose $\ _{2}n_{k}=0$ and $\
_{1}n_{k}=\partial _{k}n$ with a function $n=n(x^{k}).$ We emphasize that $%
(\partial _{i}-w_{i}\partial _{3})\Phi \equiv 0$ for any $\Phi (x^{k},y^{3})$
if $w_{i}$ is defined by (\ref{w1bl}). Introducing instead of $\Phi $ a new
functional $H(\Phi ),$ we obtain  $(\partial _{i}-w_{i}\partial _{3})H=\frac{%
\partial H}{\partial \Phi }(\partial _{i}-w_{i}\partial _{3})\Phi =0$.
Using formulas (\ref{h4bl}) for functionals of type $h_{4}=H(|\tilde{\Phi}%
(\Phi )|),$ we solve always the equations $(\partial _{i}-w_{i}\partial
_{3})h_{4}=0,$ which is equivalent to the second system of equations in (\ref%
{lccondb}) because  $(\partial _{i}-w_{i}\partial _{3})\ln \sqrt{|h_{4}|}%
\sim (\partial _{i}-w_{i}\partial _{3})h_{4}$.  For a subclass of generating
functions $\Phi =\check{\Phi}$ for which
\begin{equation}
(\partial _{i}\check{\Phi})^{\ast }=\partial _{i}\check{\Phi}^{\ast },
\label{aux4a}
\end{equation}%
we compute for the left part of the second equation in (\ref{lccondb}), $%
(\partial _{i}-w_{i}\partial _{3})\ln \sqrt{|h_{4}|}=0.$ The first system of
equations in (\ref{lccondb}) can be solved in explicit form if $w_{i}$ are
determined by formulas (\ref{w1bl}), and $h_{3}[\tilde{\Phi}]$ and $h_{4}[%
\tilde{\Phi},\tilde{\Phi}^{\ast }]$ are chosen respectively for $\Upsilon
=\Lambda .$ We can consider $\tilde{\Phi}=\tilde{\Phi}(\ln \sqrt{|h_{3}|})$
for a functional dependence $h_{3}[\tilde{\Phi}[\check{\Phi}]].$ This allows
us to obtain the formulas  $w_{i}=\partial _{i}|\tilde{\Phi}|/|\tilde{\Phi}%
|^{\ast }=\partial _{i}|\ln \sqrt{|h_{3}|}|/|\ln \sqrt{|h_{3}|}|^{\ast }$.
Taking derivative $\partial _{3}$ on both sides of this equation, we get
\begin{equation*}
w_{i}^{\ast }=\frac{(\partial _{i}|\ln \sqrt{|h_{3}|}|)^{\ast }}{|\ln \sqrt{%
|h_{3}|}|^{\ast }}-w_{i}\frac{|\ln \sqrt{|h_{3}|}|^{\ast \ast }}{|\ln \sqrt{%
|h_{3}|}|^{\ast }}.
\end{equation*}%
If the conditions (\ref{aux4a}) are satisfied, we can construct generic
off--diagonal configurations with $w_{i}^{\ast }=(\partial
_{i}-w_{i}\partial _{3})\ln \sqrt{|h_{3}|}$ which is necessary for the zero
torsion conditions. Finally, we note that the conditions $\partial
_{k}w_{i}=\partial _{i}w_{k}$ from the second line in (\ref{lccondb}) $\ $%
are solved for any
\begin{equation}
\check{w}_{i}=\partial _{i}\check{\Phi}/\check{\Phi}^{\ast }=\partial _{i}%
\widetilde{A},  \label{w1c}
\end{equation}%
with a nontrivial function $\widetilde{A}(x^{k},y^{3})$ depending
functionally on generating function $\check{\Phi}.$
\end{enumerate}

The class of off--diagonal metrics of type (\ref{dm}) constructed following
above steps 1-5 for $\Upsilon =\breve{\Upsilon}=\Lambda ,$ $\Phi =\check{\Phi%
}=\tilde{\Phi}$ and $\ _{2}n_{k}=0$ in (\ref{n1b}) are determined by
quadratic elements of type {\small
\begin{equation}
ds^{2} = e^{\psi (x^{k})}[(dx^{1})^{2}+(dx^{2})^{2}]+ \frac{(\check{\Phi}%
^{\ast })^{2}}{\Lambda \check{\Phi}^{2}}[dy^{3}+(\partial _{i}\widetilde{A}[%
\check{\Phi}])dx^{i}]^{2}-\frac{\check{\Phi}^{2}}{4|\Lambda |}[dt+(\partial
_{k}n)dx^{k}]^{2}.  \label{qelgen}
\end{equation}%
} We can consider arbitrary generating functions but take the effective
cosmological constant $\Lambda $ for a model of $f$--modified massive
gravity for a source (\ref{source1a}). If $\Upsilon =~\Lambda $ (\ref{w1bl})
is for a source (\ref{source}), we obtain an effective pseudo--Riemannian
metric with N--adapted coefficients determined by effective sources in
modified gravity. Via nonlinear off--diagonal interactions in GR,
corresponding certain effective sources encoding contributions from modified
gravity, we mimic both massive gravitational and/or $f$--functional
contributions. Here we emphasize that off--diagonal configurations (of
vacuum and non--vacuum type) are possible even the effective sources from
modified bi--metric gravity are constrained to be zero.

For arbitrary $\phi$ and $\Upsilon ,$ and related $\Phi ,$ or $\widetilde{%
\Phi },$ and $\Lambda ,$ we can generate off--diagonal solutions of (\ref%
{eq1b})--(\ref{eq4b}) with nonholonomically induced torsion, {\small
\begin{eqnarray}
ds^{2} &=&e^{\psi (x^{k})}[(dx^{1})^{2}+(dx^{2})^{2}]+  \label{qelgent} \\
&&(z_{3})^{2}[dy^{3}+\frac{\partial _{i}\Phi }{\Phi ^{\ast }}%
dx^{i}]^{2}-(z_{4})^{2}[dt+(\ _{1}n_{k}+\ _{2}n_{k}\int dy^{3}\frac{%
(z_{3})^{2}}{(z_{4})^{3}}dx^{k})]^{2},  \notag
\end{eqnarray}%
} for $\epsilon _{3}=1,\epsilon _{4}=-1,$ where the functions $%
z_{3}(x^{k},y^{3})$ and $z_{4}(x^{k},y^{3})$ are defined by formulas (\ref%
{h3a}) and (\ref{h4a}). In N--adapted frames, the ansatz for such solutions
define a nontrivial distorting tensor as in $\widehat{\mathbf{Z}}=\{\widehat{%
\mathbf{Z}}_{\ \beta \gamma }^{\alpha }\}$ in (\ref{distr}).

\subsection{Formal integration via polarization functions}

We can not distinguish the coefficients and multiples in a general
off--diagonal solution (\ref{qelgen}) and (\ref{qelgent}) which are
determined by a prime fiducial, $f$--modified and/or any diagonal exact
solution in GR. Such contributions mix for general coordinate/frame
transforms. Our goal is to find certain parameterizations of target metrics
when the coefficients of prime metrics can be defined in explicit form
together with possible "gravitational polarizations" of effective constants
and nonholonomic deformations of the coefficients of metrics. For certain
additional assumptions, such deformations can be parameterized on a small
parameter.

\subsubsection{Levi--Civita deformations in massive gravity}

Metrics of type (\ref{dm}) can be used for constructing nonholonomic
deformations $(\mathbf{\mathring{g}},\mathbf{\mathring{N},\ }^{v}\mathring{%
\Upsilon},\mathring{\Upsilon})\rightarrow (\widetilde{\mathbf{g}},\widetilde{%
\mathbf{N}}\mathbf{,\ }^{v}\widetilde{\Upsilon },\widetilde{\Upsilon }),$
when the prime metric $\mathbf{\mathring{g}}$ may be, or not, an exact
solution of the Einstein or other modified gravitational equations but the
target metric $\mathbf{g}$ positively defines a generic off--diagonal
solution of field equations in a model of gravity.

We are interested in deformations of metrics $\mathbf{\mathring{g}(}x^{k}%
\mathbf{)}$ possessing two Killing vector symmetries (in particular, such a
metric may define a black hole, or wormhole solution). The N--adapted
deformations of coefficients of metrics, frames and sources are chosen in
the form {\small
\begin{eqnarray*}
[\mathring{g}_{i},\mathring{h}_{a},\mathring{w}_{i},\mathring{n}%
_{i}]\rightarrow [\widetilde{g}_{i}=\widetilde{\eta }_{i}\mathring{g}_{i},%
\widetilde{h}_{3}=\widetilde{\eta }_{3}\mathring{h}_{3},\widetilde{h}_{4}=%
\widetilde{\eta }_{4}\mathring{h}_{4},\widetilde{w}_{i}=\mathring{w}_{i}+\
^{\eta }w_{i},n_{i}=\mathring{n}_{i}+\ ^{\eta }n_{i}],  \notag \\
\ ^{v}\widetilde{\Upsilon }=\ ^{v}\hat{\Upsilon}(x^{k})\ ^{v}\mathring{%
\Upsilon},\ ^{v}\hat{\Upsilon}(x^{k})=\tilde{\Upsilon}=\mu _{g}^{2}\ \lambda
(x^{k})(\mathring{h}_{3})^{-1},\check{\Phi}^{2}=\exp [2\varpi ]\ \mathring{h}%
_{3}\mathring{h}_{4},  \label{defrel}
\end{eqnarray*}%
} where the source $\mu _{g}^{2}\ \lambda (x^{k})$ for massive gravity is
taken as in (\ref{source1a}) and the \ values \ $\widetilde{\eta }_{a},%
\widetilde{w}_{i},\tilde{n}_{i}$ and $\varpi $ are functions on three
coordinates $(x^{k},y^{3})$ and $\widetilde{\eta }_{i}(x^{k})$ depends only
on h--coordinates. The prime data $\mathring{g}_{i},\mathring{h}_{a},%
\mathring{w}_{i},\mathring{n}_{i}$ (which can be determined by an exact
solution in gravity theory, by any fiducial metric) are given by
coefficients depending only on $(x^{k}).$ The value $\ ^{v}\mathring{\Upsilon%
}$ can be defined from certain physical assumptions on matter and effective
sources if $\mathbf{\mathring{g}}$ chosen as a solution of certain
gravitational field equations in a theory of gravity. Conventionally, we can
take $\ ^{v}\mathring{\Upsilon}=1$ if, for instance, a general
pseudo--Riemannian metric $\mathbf{\mathring{g}}$ is transformed into a
solution of some (generalized) field equations with source $(\mathbf{\ }^{v}%
\widetilde{\Upsilon },\widetilde{\Upsilon }).$

In terms of $\eta $--functions resulting in $h_{a}^{\ast }\neq 0$ and $%
g_{i}=c_{i}e^{\psi {(x^{k})}},$ the solutions (\ref{qelgen}) can be
re--written in the form%
\begin{eqnarray}
ds^{2} &=&e^{\psi (x^{k})}[(dx^{1})^{2}+(dx^{2})^{2}]+  \label{nvlcmgs} \\
&&\frac{(\varpi ^{\ast })^{2}}{\ \mu _{g}^{2}\ \lambda }\mathring{h}%
_{3}[dy^{3}+(\partial _{i}\ ^{\eta }\widetilde{A})dx^{i}]^{2}-\frac{%
e^{2\varpi }}{4\mu _{g}^{2}|\ \lambda |}\mathring{h}_{4}[dt+(\partial _{k}\
^{\eta }n(x^{i}))dx^{k}]^{2}.  \notag
\end{eqnarray}%
The gravitational polarizations $(\eta _{i},\eta _{a})$ and N--coefficients $%
(w_{i},n_{i})$ are computed following formulas\
\begin{eqnarray*}
e^{\psi (x^{k})} &=&\widetilde{\eta }_{1}\mathring{g}_{1}=\widetilde{\eta }%
_{2}\mathring{g}_{2},\ \widetilde{\eta }_{3}=\frac{(\varpi ^{\ast })^{2}}{\
\mu _{g}^{2}\ \lambda },\widetilde{\eta }_{4}=\frac{e^{2\varpi }}{4\mu
_{g}^{2}|\lambda |}, \\
w_{i} &=&\mathring{w}_{i}+\ ^{\eta }w_{i}=\partial _{i}(\ ^{\eta }\widetilde{%
A}[\varpi ]),\ n_{k}=\mathring{n}_{k}+\ ^{\eta }n_{k}=\partial _{k}(\ ^{\eta
}n),
\end{eqnarray*}%
where $\ ^{\eta }\widetilde{A}(x^{k},y^{3})$ is introduced via formulas and
assumptions similar to (\ref{aux4a})--(\ref{w1c}) and $\psi ^{\bullet
\bullet }+\psi ^{\prime \prime }=2\ ^{v}\hat{\Upsilon}(x^{k})\ ^{v}\mathring{%
\Upsilon}.$ For N--coefficients, there are used the parameterizations  $%
w_{i}=\mathring{w}_{i}+\ ^{\eta }w_{i}=\partial _{i}(\ e^{\varpi }\sqrt{|%
\mathring{h}_{3}\mathring{h}_{4}|})/\ \varpi ^{\ast }e^{\varpi }\sqrt{|%
\mathring{h}_{3}\mathring{h}_{4}|}=\partial _{i}\ ^{\eta }\widetilde{A}$.
We can take any function $\ ^{\eta }n(x^{k})$ and put $\lambda =const\neq 0$
for both the prime (if this one is an exact solution with nontrivial
cosmological constant) and target metrics.

\subsubsection{Induced torsion in massive gravity}

This class of solutions with nontrivial d--torsion (\ref{dtors}) is
determined by a metric (\ref{qelgent}) when the coefficients (\ref{h4a})-- (%
\ref{n1b}) are computed for the source $\Upsilon =\ \mu _{g}^{2}\ \lambda
(x^{k})$ in massive gravity and for possible effective anisotropic
polarizations. The corresponding off--diagonal quadratic element is given by
\begin{eqnarray}
ds^{2} &=&e^{\psi (x^{k})}[(dx^{1})^{2}+(dx^{2})^{2}]+\frac{(\Phi ^{\ast
})^{2}}{\ \mu _{g}^{2}\lambda \Phi ^{2}}[dy^{3}+\frac{\partial _{i}\Phi }{%
\Phi ^{\ast }}dx^{i}]^{2}  \notag \\
&&-\frac{\Phi ^{2}}{4\mu _{g}^{2}|\lambda |}[dt+(\ _{1}n_{k}+\ _{2}n_{k}%
\frac{4\mu _{g}(\Phi ^{\ast })^{2}}{\Phi ^{5}})dx^{k}]^{2}.  \label{ofindtmg}
\end{eqnarray}%
We can see that nontrivial stationary off--diagonal torsion effects may
result in additional effective rotation proportional to $\mu _{g}$ if the
integration function $\ _{2}n_{k}\neq 0.$ Such terms do not exist for the
Levi--Civita massive configurations of type (\ref{nvlcmgs}). Using different
classes of off--diagonal metrics (\ref{ofindtmg}) and (\ref{nvlcmgs}) we can
study if a massive gravity theory is with induced torsion or characterized
by additional nonholonomic constraints as GR and zero torsion.

\subsubsection{Small $f$--modifications and massive gravity}

Additional modifications of GR are possible by $f$--functionals with an
effective source $\Lambda $ (\ref{source1a}). Using two nonholonomic
deformations  $(\mathbf{\mathring{g}},\mathbf{\mathring{N},\ }^{v}\mathring{%
\Upsilon},\mathring{\Upsilon})\rightarrow (\widetilde{\mathbf{g}},\widetilde{%
\mathbf{N}}\mathbf{,\ }^{v}\widetilde{\Upsilon },\widetilde{\Upsilon }%
)\rightarrow (\mathbf{g}[\varepsilon ],\mathbf{N}[\varepsilon ]\mathbf{,\ }%
\Lambda )$,  we construct off--diagonal solutions type (\ref{dm}) with $%
\mathbf{g}$ and$\ \mathbf{N}$ depending on a small parameter $\varepsilon ,$
$0<\varepsilon \ll 1,$ when the source in massive gravity $\mu
_{g}^{2}|\lambda |$ is generalized to an effective cosmological constant $%
\Lambda $ with additional contributions by matter fields and $f$%
--modifications of gravity. The corresponding N--adapted transforms are
parameterized {\small
\begin{eqnarray}
&&[\mathring{g}_{i},\mathring{h}_{a},\mathring{w}_{i},\mathring{n}%
_{i}]\rightarrow \lbrack g_{i}=(1+\varepsilon \chi _{i})\widetilde{\eta }_{i}%
\mathring{g}_{i},h_{3}=(1+\varepsilon \chi _{3})\widetilde{\eta }_{3}%
\mathring{h}_{3},h_{4}=(1+\varepsilon \chi _{4})\widetilde{\eta }_{4}%
\mathring{h}_{4},  \notag \\
&&\ ^{\varepsilon }w_{i} =\mathring{w}_{i}+\widetilde{w}_{i}+\varepsilon
\overline{w}_{i},\ ^{\varepsilon }n_{i}=\mathring{n}_{i}+\tilde{n}%
_{i}+\varepsilon \overline{n}_{i}],  \label{def2} \\
&&\ \mu _{g}^{2}\ \lambda (x^{k})=\Lambda \lbrack 1-\varepsilon \ ^{\mu
}\chi (x^{k})],\varpi ^{\ast }[\varepsilon ]=\varpi ^{\ast }(1+\varepsilon \
^{\varpi }\chi (x^{k},y^{3})),  \notag
\end{eqnarray}%
} where the \ values $\chi _{i}(x^{k}),\ ^{\lambda }\chi (x^{k}),\overline{n}%
_{i}(x^{k}),$ $\ ^{\varpi }\chi (x^{k},y^{3}),\chi _{a}(x^{k},y^{3})$ and $%
\overline{w}_{i}$\ $(x^{k},y^{3})$ can be computed to define
LC--configurations as solutions of the system (\ref{eq1b})--(\ref{lccondb}).

The deformations (\ref{def2}) of the off--diagonal solutions (\ref{nvlcmgs})
result in a new class of $\varepsilon $--deformed solutions if%
\begin{eqnarray*}
\chi _{3} &=&\ ^{\mu }\chi +\ ^{\varpi }\chi ,\chi _{4}=\ ^{\mu }\chi
+\varpi ^{-1}\int dy^{3}(\ ^{\varpi }\chi \varpi ^{\ast }), \\
\overline{w}_{i} &=&\partial _{i}(\ \ ^{\varpi }\chi \sqrt{|\mathring{h}_{3}%
\mathring{h}_{4}|})/\ \varpi ^{\ast }e^{\varpi }\sqrt{|\mathring{h}_{3}%
\mathring{h}_{4}|}=\partial _{i}\overline{A},\overline{n}_{i}=\partial _{i}%
\overline{n}.
\end{eqnarray*}%
The coefficients for the h--metric $g_{i}=\exp \psi (x^{i})=(1+\varepsilon
\chi _{i})\widetilde{\eta }_{i}\mathring{g}_{i}$ are solutions of (\ref{eq1b}%
) with $\mathbf{\ }^{v}\Upsilon =\Lambda =\breve{\Upsilon}(x^{k})+\mu
_{g}^{2}\lambda ,$ where $\breve{\Upsilon}(x^{k})$ is determined by possible
contributions of matter fields and $f$--modifications parametrized in (\ref%
{source1a}).

In next sections, we shall construct such solutions in explicit form for
ellipsoid, toroid and solitonic deformations. \ If $\varepsilon $%
--deformations of type (\ref{def2}) are considered for metrics (\ref%
{ofindtmg}), we can generate new classes of off--diagonal solutions with
nonholonomically induced torsion determined both by massive and $f$%
--modifications of GR.

\section{Off--diagonal Deformations of Wormhole Metrics}

\label{s4} In this section, we construct and analyze two examples when a
wormhole solution matching an exterior Schwarzschild -- de Sitter spacetime
is nonholonomically deformed into new classes of off--diagonal solutions.
The target metrics are constructed for modifications of GR with effectively
polarized cosmological constants and "polarization" multiples and additional
terms to, respectively, diagonal and non--diagonal coefficients of metrics.
The deformations resulting from massive gravity are studied for an
effectively polarized cosmological constant proportional to $\mu _{g}^{2}.$
Modifications determined by $f$--terms are computed for a small deformation
parameter $\varepsilon .$

\subsection{Prime metrics for 4--d wormholes}

Let us consider a diagonal prime wormhole metric
\begin{eqnarray}
\mathbf{\mathring{g}} &=&\mathring{g}_{i}(x^{k})dx^{i}\otimes dx^{i}+%
\mathring{h}_{a}(x^{k})dy^{a}\otimes dy^{a}  \label{whm} \\
&=&[1-b(r)/r]^{-1}dr\otimes dr+r^{2}(d\theta \otimes d\theta +\sin
^{2}\theta d\varphi \otimes d\varphi )-e^{2B(r)}dt\otimes dt,  \notag
\end{eqnarray}%
where $B(r)$ and $b(r)$ are called respectively the red--shift and form
functions, see details in \cite{morris,visser,lobo,boehm}. The radial
coordinate has a range $r_{0}\leq r<a,$ where the minimum value $r_{0}$ is
for the wormhole throat and $a$ is the distance at which the interior
spacetime joins to an exterior vacuum solution ($a\rightarrow \infty $ for
specific asymptotically flat wormhole geometries). Certain conditions have
to be imposed on coefficients of (\ref{whm}) and on diagonal components of
the stress--energy tensor
\begin{equation}
\mathring{T}_{~\nu }^{\mu }=diag[~^{r}p=\tau (r),~^{\theta
}p=p(r),~^{\varphi }p=p(r),~^{t}p=\rho (r)]  \label{wemt}
\end{equation}%
in order to generate wormhole solutions of the Einstein equations in GR.

A well known class of wormhole metrics is constructed to possess conformal
symmetry determined \ by a vector $\mathbf{X}=\{X^{\alpha }(u)\},$ when the
Lie derivative $X^{\alpha }\partial _{\alpha }\mathring{g}_{\mu \nu }+%
\mathring{g}_{\alpha \nu }\partial _{\mu }X^{\alpha }+\mathring{g}_{\alpha
\mu }\partial _{\nu }X^{\alpha }=\sigma \mathring{g}_{\mu \nu }$, where $%
\sigma = \sigma (u)$ is the conformal factor. Such solutions are
parameterized by {\small
\begin{eqnarray}
B(r) &=&\frac{1}{2}\ln (C^{2}r^{2})-\kappa \int r^{-1}\left( 1-b(r)/r\right)
^{-1/2}dr,~b(r)=r[1-\sigma ^{2}(r)],  \notag \\
\tau (r) &=&\frac{1}{\kappa ^{2}r^{2}}(3\sigma ^{2}-2\kappa \sigma -1),~p(r)=%
\frac{1}{\kappa ^{2}r^{2}}(\sigma ^{2}-2\kappa \sigma +\kappa ^{2}+2r\sigma
\sigma ^{\bullet }),  \notag \\
\rho (r) &=&\frac{1}{\kappa ^{2}r^{2}}(1-\sigma ^{2}-2r\sigma \sigma
^{\bullet }).  \label{data1}
\end{eqnarray}%
} The data (\ref{data1}) generate \textquotedblright
diagonal\textquotedblright\ wormhole configurations determined by
\textquotedblright exotic\textquotedblright\ matter because the null energy
condition (NEC), $\mathring{T}_{\mu \nu }k^{\mu }k^{\nu }\geq 0$, ($k^{\nu }$
is any null vector), is violated.

We shall study configurations which match the interior geometries to an
exterior de Sitter one which (in general) can be also determined by an
off--diagonal metric. The exotic matter and effective matter configurations
are considered to be restricted to spacial distributions in the throat
neighborhood which limit the dimension of locally isotropic and/or
anisotropic wormhole to be not arbitrarily large.

\subsection{Parametric deformations and exterior de Sitter spacetimes}

The Schwarzschild -- de Sitter (SdS) metric%
\begin{equation}
ds^{2}=q^{-1}(r)(dr^{2}+r^{2}\ d\theta ^{2})+r^{2}\sin ^{2}\theta \ d\varphi
^{2}-q(r)\ dt^{2},  \label{sds}
\end{equation}%
can be re--parameterized for any $(x^{1}(r,\theta ),x^{2}(r,\theta
),y^{3}=\varphi ,y^{4}=t)$ when
\begin{equation*}
q^{-1}(r)(dr^{2}+r^{2}\ d\theta ^{2})=e^{\mathring{\psi}{(x^{k})}%
}[(dx^{1})^{2}+(dx^{2})^{2}].
\end{equation*}
Such a metric defines two real static solutions of the Einstein equations
with cosmological constant $\Lambda $ if $M<1/3\sqrt{|\Lambda |},$ for $%
q(r)=1-2\overline{M}(r)/r,\overline{M}(r)=M+\Lambda r^{3}/6,$ where $M$ is a
constant mass parameter. For diagonal configurations, we can identify $%
\Lambda $ with the effective cosmological constant (\ref{source1a}).

In this work, we study conformal, ellipsoid and/or solitonic/toroidal
deformations related in certain limits to the Schwarzschild -- de Sitter
metric written in the form
\begin{equation}
~_{\Lambda }\mathbf{g}=d\xi \otimes d\xi +r^{2}(\xi )\ d\theta \otimes
d\theta +r^{2}(\xi )\sin ^{2}\theta \ d\varphi \otimes d\varphi -q(\xi )\
dt\otimes \ dt,  \label{sds1}
\end{equation}%
for local coordinates
\begin{equation}
x^{1}=\xi =\int dr/\sqrt{\left\vert q(r)\right\vert },x^{2}=\vartheta
,y^{3}=\varphi ,y^{4}=t,  \label{coordsph}
\end{equation}
for a system of $h$--coordinates when $(r,\theta )\rightarrow (\xi
,\vartheta )$ with $\xi $ and $\vartheta $ of length dimension. The data for
this primary metric are written {\small
\begin{equation*}
\mathring{g}_{i}=\mathring{g}_{i}(x^{k})=e^{\mathring{\psi}{(x^{k})}},\
\mathring{h}_{3}=r^{2}(x^{k})\sin ^{2}\theta (x^{k}),\mathring{h}_{4}=-q(r{%
(x^{k})}),\mathring{w}_{i}=0,\ \mathring{n}_{i}=0.
\end{equation*}%
} Let us analyze how such diagonal metrics can be off--diagonally deformed
by contributions from massive and $f$--modified gravity:

\subsubsection{Off--diagonal de Sitter deformations in massive gravity}

Solutions resulting in the Levi--Civita configurations can be generated
similarly to (\ref{nvlcmgs}) but using data (\ref{sds1}) {\small
\begin{eqnarray}
&&ds^{2} =e^{\tilde{\psi}(\xi ,\theta )}(d\xi ^{2}+\ d\vartheta ^{2})+ \frac{%
(\varpi ^{\ast })^{2}}{\ \mu _{g}^{2}\ \lambda (\xi ,\vartheta )}r^{2}(\xi
)\sin ^{2}\theta (\xi ,\vartheta ) [d\varphi +(\partial _{\xi }\ ^{\eta }%
\widetilde{A})d\xi +  \notag \\
&& (\partial _{\vartheta }\ ^{\eta }\widetilde{A})d\vartheta ]^{2}- \frac{%
e^{2\varpi }}{4\mu _{g}^{2}|\ \lambda (\xi ,\vartheta )|}q(\xi )[dt+\partial
_{\xi }\ ^{\eta }n(\xi ,\vartheta )d\xi +\partial _{\vartheta }\ ^{\eta
}n(\xi ,\vartheta )d\vartheta ]^{2},  \label{masds}
\end{eqnarray}%
} where $e^{\tilde{\psi}(\xi ,\vartheta )}=\widetilde{\eta }_{1}\mathring{g}%
_{1}=\widetilde{\eta }_{2}\mathring{g}_{2}$ are solutions of $\tilde{\psi}%
^{\bullet \bullet }+\tilde{\psi}^{\prime \prime }=2\ \mu _{g}^{2}\ \lambda
(\xi ,\vartheta ).$ The generating function $\varpi (\xi ,\vartheta ,\varphi
),$ effective source $\lambda (\xi ,\vartheta )$ and mass parameter $\mu
_{g} $ should be fixed from physical assumptions on systems of reference,
fixed prime St\"{u}kelberg fields (using algebraic conditions of type(\ref%
{stuk})) and observable effects in modern cosmology. The value $\widetilde{n}%
_{i}=\ ^{\eta }n_{i}(\xi ,\vartheta )=\partial _{i}\ ^{\eta }n(\xi
,\vartheta )$ is an integration function and $^{\eta }\widetilde{A}(\xi
,\vartheta ,\varphi )$ is determined by $e^{2\varpi }$ following formula (%
\ref{w1c}) and
\begin{equation}
\widetilde{w}_{i}=\ ^{\eta }w_{i}=\frac{\partial _{i}(\ e^{\varpi }r(\xi
)\sin \theta (\xi ,\vartheta )\sqrt{|q(\xi )|})}{\varpi ^{\ast }e^{\varpi
}r(\xi )\sin \theta (\xi ,\vartheta )\sqrt{|q(\xi )|}}\ =\partial _{i}\
^{\eta }\widetilde{A},\mbox{ for }x^{i}=(\xi ,\vartheta ).  \label{wcoef}
\end{equation}

It should be noted here that the N--coefficients in (\ref{masds}) result in
nonzero anholonomy coefficients (\ref{anhcoef}) for nonolonomic relations of
type (\ref{nonholr}). This proves that such solutions can not be
diagonalized via frame/coordinate transforms and that, in general, they are
characterized by six (from possible ten) independent coefficients of
metrics. We can mimic such configurations by off--diagonal interactions in
GR with corresponding effective matter source determined by terms induced by
$\mu _{g}$ taken as an integration parameter. It can be related to Killing
symmetries of such metrics, see details in Ref. \cite{vparam}.

\subsubsection{Ellipsoidal $f$--modifications}

Deformations (\ref{def2}) on a parameter $\varepsilon ,$ $0\leq \varepsilon
<1,$ are considered for the solutions in massive gravity (\ref{masds}), with
\begin{eqnarray}
\chi _{3} &=&\ ^{\varpi }\chi ,\chi _{4}=\varpi ^{-1}\int d\varphi (\
^{\varpi }\chi \varpi ^{\ast }),  \label{def2a} \\
\overline{w}_{i} &=&\frac{\partial _{i}(\ \ ^{\varpi }\chi r(\xi )\sin
\theta (\xi ,\vartheta )\sqrt{|q(\xi )|})}{\varpi ^{\ast }e^{\varpi }r(\xi
)\sin \theta (\xi ,\vartheta )\sqrt{|q(\xi )|}}\ =\partial _{i}\overline{A},%
\overline{n}_{i}=\partial _{i}\overline{n},  \notag
\end{eqnarray}%
for $x^{i}=(\xi ,\vartheta )$ and we fix, for simplicity, $\ ^{\mu }\chi =0$
(a possible physical motivation is to consider models with constant mass
gravity parameter and zero related polarization). The coefficients of the
h--metric $g_{i}=\exp \psi (\xi ,\vartheta )=(1+\varepsilon \chi _{i})%
\widetilde{\eta }_{i}\mathring{g}_{i}$ are solutions of (\ref{eq1b}) with $\
^{v}\Upsilon =\Lambda =\breve{\Upsilon}(x^{k})+\mu _{g}^{2}\lambda ,$ where $%
\breve{\Upsilon}(\xi ,\vartheta )$ is determined by possible contributions
of matter fields and $f $--modifications parameterized in (\ref{source1a}).
The resulting target off--diagonal quadratic element is parameterized in the
form {\small
\begin{eqnarray*}
\mathbf{ds}^{2} &=&e^{\tilde{\psi}(\xi ,\vartheta )}\left( d\xi ^{2}+\
d\vartheta ^{2}\right) +\frac{(\varpi ^{\ast })^{2}}{\ \mu _{g}^{2}\ \lambda
(\xi ,\vartheta )}[1+\varepsilon \chi _{3}(\xi ,\vartheta ,\varphi
)]r^{2}(\xi )\sin ^{2}\theta (\xi ,\vartheta )(\delta \varphi )^{2} \\
&&-\frac{e^{2\varpi }}{4\mu _{g}^{2}|\ \lambda (\xi ,\vartheta )|}%
[1+\varepsilon \chi _{4}(\xi ,\vartheta ,\varphi )]q(\xi )(\delta t)^{2}, \\
\delta \varphi &=&d\varphi +[\widetilde{w}_{i}(\xi ,\vartheta ,\varphi
)+\varepsilon \overline{w}_{i}(\xi ,\vartheta ,\varphi )]dx^{i},\ \delta t =
dt+\ [\tilde{n}_{i}(\xi ,\vartheta )+\varepsilon \overline{n}_{i}(\xi
,\vartheta )]dx^{i},\
\end{eqnarray*}%
} when $\widetilde{w}_{i}$ are given by formulas (\ref{wcoef}). For such
small deformations re--parameterized in $(r,\theta )$--coordinates, the
coefficient
\begin{equation}
h_{4}=-\frac{e^{2\varpi }}{4\mu _{g}^{2}|\ \lambda (\xi ,\vartheta )|}%
[1+\varepsilon \chi _{4}]q\simeq -\frac{e^{2\varpi }}{4\mu _{g}^{2}|\
\lambda (r,\theta )|}[1-\frac{2M(r,\theta ,\varphi )}{r}]  \label{defhor}
\end{equation}%
is related to small gravitational polarizations of mass coefficients,
\begin{equation*}
M(r,\theta ,\varphi )\simeq \overline{M}(r)[1+\varepsilon (1-\frac{r}{2%
\overline{M}})\chi _{4}(r,\theta ,\varphi )].
\end{equation*}

We generate rotoid $f$--deformations if
\begin{equation}
\chi _{4}=\overline{\chi }_{4}(r,\varphi ):=\frac{2\overline{M}(r)}{r}\left(
1-\frac{2\overline{M}(r)}{r}\right) ^{-1}\underline{\zeta }\sin (\omega
_{0}\varphi +\varphi _{0}),  \label{rotchi}
\end{equation}%
for some constants $\underline{\zeta },\omega _{0}$ and $\varphi _{0},$
taken as a polarization function. With respect to N--adapted frames, there
is a smaller \textquotedblright ellipsoidal horizon\textquotedblright\ (when
$\ h_{4}=0$ in (\ref{defhor}), we get the parametric equation for an
ellipse),
\begin{equation*}
\ r_{+}\simeq \frac{2\ \overline{M}(\ r_{+})}{1+\varepsilon \underline{\zeta
}\sin (\omega _{0}\varphi +\varphi _{0})},
\end{equation*}%
where $\varepsilon $ is the eccentricity parameter. Using formulas (\ref%
{def2a}) for a prescribed value $\varpi (r,\theta ,\varphi ), $ and $\chi
_{3}=\ ^{\varpi }\chi =\partial _{\varphi }[\overline{\chi }_{4}\varpi
]/\partial _{\varphi }\varpi$, we compute
\begin{equation*}
\overline{w}_{i}=\frac{\partial _{i}(\ r(\xi )\sin \theta (\xi ,\vartheta )%
\sqrt{|q(\xi )|}\partial _{\varphi }[\overline{\chi }_{4}\varpi ])}{%
e^{\varpi }r(\xi )\sin \theta (\xi ,\vartheta )\sqrt{|q(\xi )|}\partial
_{\varphi }\varpi }=\partial _{i}\overline{A}.
\end{equation*}

The resulting solutions in massive $f$--gravity with rotoid symmetry are
parameterized {\small
\begin{eqnarray}
~\mathbf{ds}^{2} &=&e^{\tilde{\psi}(\xi ,\vartheta )}\left( d\xi ^{2}+\
d\vartheta ^{2}\right) +\frac{(\varpi ^{\ast })^{2}}{\ \mu _{g}^{2}\ \lambda
}(1+\varepsilon \frac{\partial _{\varphi }[\overline{\chi }_{4}\varpi ]}{%
\partial _{\varphi }\varpi })r^{2}(\xi )\sin ^{2}\theta (\xi ,\vartheta )\
(\delta \varphi )^{2}  \notag \\
&&-\frac{e^{2\varpi }}{4\mu _{g}^{2}|\ \lambda |}[1+\varepsilon \overline{%
\chi }_{4}]q(\xi )\ (\delta t)^{2},  \label{massivsmall} \\
\delta \varphi &=&d\varphi +[\partial _{i}\ ^{\eta }\widetilde{A}%
+\varepsilon \partial _{i}\overline{A}]dx^{i},\ \delta t=dt+[\partial _{i}\
^{\eta }n+\varepsilon \partial _{i}\overline{n}]~dx^{i}.  \notag
\end{eqnarray}%
} Such stationary configurations are generated by nonlinear off--diagonal
interactions in massive gravity with nontrivial $\mu _{g}^{2}\ \lambda (\xi
,\vartheta )$ terms. We note that, in general, the limit $\mu
_{g}\rightarrow 0$ is not smooth for such classes of solutions. There are
necessary additional assumptions on nonholonomic constraints resulting in
diagonal metrics with two--Killing symmetries or for selecting black rotoid
-- de Sitter configurations. It is possible to model such solutions via
locally anisotropic effective polarizations of the coefficients of metrics
and physical constants (treated as integration functions and constants) in
GR. For this class of solutions, the contributions related to "massive"
gravity terms are very different from those generated by $f$--deformations.
In the last case, there are smooth limits for $\varepsilon \rightarrow 0,$
when (for instance) rotoid symmetries may transform into spherical ones.

\section{Ellipsoid, Solitonic \& Toroid Deformations of \newline
Wormhole Metrics}

\label{s5} In this section, we explore rotoid deformations of wormhole
configurations determined by off--diagonal effects in massive gravity and $f$%
--modificati\-ons. The general ansatz for such metrics is taken in the form
{\small
\begin{eqnarray}
\mathbf{ds}^{2} &=&e^{\tilde{\psi}(\widetilde{\xi },\theta )}(d\widetilde{%
\xi }^{2}+\ d\vartheta ^{2})+  \label{offdwans1} \\
&&\frac{\lbrack \partial _{\varphi }\varpi (\widetilde{\xi },\vartheta
,\varphi )]^{2}}{\ \mu _{g}^{2}\ \lambda (\widetilde{\xi },\vartheta )}%
\left( 1+\varepsilon \frac{\partial _{\varphi }[\chi _{4}(\widetilde{\xi }%
,\vartheta ,\varphi )\varpi (\widetilde{\xi },\vartheta ,\varphi )]}{%
\partial _{\varphi }\varpi (\widetilde{\xi },\vartheta ,\varphi )}\right)
r^{2}(\widetilde{\xi })\sin ^{2}\theta (\widetilde{\xi },\vartheta )(\delta
\varphi )^{2}  \notag \\
&&-\frac{e^{2\varpi (\widetilde{\xi },\vartheta ,\varphi )]}}{4\mu
_{g}^{2}|\ \lambda (\widetilde{\xi },\vartheta )|}[1+\varepsilon \chi _{4}(%
\widetilde{\xi },\vartheta ,\varphi )]e^{2B(\widetilde{\xi })}(\delta t)^{2},
\notag \\
\delta \varphi &=&d\varphi +\partial _{\widetilde{\xi }}[\ ^{\eta }%
\widetilde{A}(\widetilde{\xi },\vartheta ,\varphi )+\varepsilon \overline{A}(%
\widetilde{\xi },\vartheta ,\varphi )]d\widetilde{\xi }+\partial _{\vartheta
}[\ ^{\eta }\widetilde{A}(\widetilde{\xi },\vartheta ,\varphi )+\varepsilon
\overline{A}(\widetilde{\xi },\vartheta ,\varphi )]d\vartheta ,\   \notag \\
\delta t &=&dt+\partial _{\widetilde{\xi }}[\ ^{\eta }n(\widetilde{\xi }%
,\vartheta )+\varepsilon \partial _{i}\overline{n}(\widetilde{\xi }%
,\vartheta )]~d\widetilde{\xi }+\partial _{\vartheta }[\ ^{\eta }n(%
\widetilde{\xi },\vartheta )+\varepsilon \partial _{i}\overline{n}(%
\widetilde{\xi },\vartheta )]~d\vartheta ,  \notag
\end{eqnarray}%
} where $\ \widetilde{\xi }=\int dr/\sqrt{|1-b(r)/r|}$ and $B(\widetilde{\xi
}) $ are determined by the prime metric (\ref{whm}). We can chose such
generating and integration functions when the metrics (in corresponding
limits) define exterior spacetimes (\ref{massivsmall}), for coordinates (\ref%
{coordsph}) and $e^{2B(\widetilde{\xi })}\rightarrow q(r),$ see (\ref{sds1}).

The class of solutions (\ref{offdwans1}) are for stationary configurations
determined by respective general and small $\varepsilon $--parametric $\mu
_{g}$-- and $f$--modifications.

\subsection{Ellipsoidal off--diagonal wormhole deformations}

Rotoid $\varepsilon $--configurations are "extracted" from (\ref{offdwans1})
if we take for the $f$--deformations {\small
\begin{equation}
\chi _{4}=\overline{\chi }_{4}(r,\varphi ):=\frac{2\overline{M}(r)}{r}\left(
1-\frac{2\overline{M}(r)}{r}\right) ^{-1}\underline{\zeta }\sin (\omega
_{0}\varphi +\varphi _{0}),  \label{def2b}
\end{equation}%
} for $r$ considered as a function $r(\ \widetilde{\xi })$. This is
different from $r(\xi )$ taken in previous section but may be parameterized
to have $r(\ \widetilde{\xi })\rightarrow r(\xi )$ in exterior spacetimes.
Let us define {\small
\begin{equation*}
h_{3} = \widetilde{\eta }_{3}(\widetilde{\xi },\vartheta ,\varphi )\left[
1+\varepsilon \chi _{3}(\widetilde{\xi },\vartheta ,\varphi )\right] \ ^{0}%
\overline{h}_{3}(\widetilde{\xi },\vartheta ),\ h_{4} = \widetilde{\eta }%
_{4}(\widetilde{\xi },\vartheta ,\varphi )\left[ 1+\varepsilon \overline{%
\chi }_{4}(\widetilde{\xi },\varphi )\right] \ ^{0}\overline{h}_{4}(%
\widetilde{\xi }),
\end{equation*}%
} for $\ ^{0}\overline{h}_{3}=r^{2}(\widetilde{\xi })\sin ^{2}\theta (%
\widetilde{\xi },\vartheta ),$ $\ ^{0}\overline{h}_{4}=q(\widetilde{\xi })$
and%
\begin{equation}
\widetilde{\eta }_{3}=\frac{[\partial _{\varphi }\varpi (\widetilde{\xi }%
,\vartheta ,\varphi )]^{2}}{\ \mu _{g}^{2}\ \lambda (\widetilde{\xi }%
,\vartheta )},\widetilde{\eta }_{4}=\frac{e^{2\varpi (\widetilde{\xi }%
,\vartheta ,\varphi )]}}{4\mu _{g}^{2}|\ \lambda (\widetilde{\xi },\vartheta
)|q(\widetilde{\xi })}e^{2B(\widetilde{\xi })},  \label{polfew}
\end{equation}%
where $e^{2B(\widetilde{\xi })}\rightarrow q(\widetilde{\xi })$ if $%
\widetilde{\xi }\rightarrow \xi .$ Introducing (\ref{def2b}) in respective
formulas (\ref{def2a}) for any prescribed generating function $\widetilde{%
\varpi }(\widetilde{\xi },\vartheta ,\varphi ),$ we can compute
\begin{eqnarray*}
&& \widetilde{\chi }_{3}=\chi _{3}(\widetilde{\xi },\vartheta ,\varphi )=\
^{\varpi }\chi =\partial _{\varphi }[\overline{\chi }_{4}\widetilde{\varpi }%
]/\partial _{\varphi }\widetilde{\varpi }, \mbox{ and } \\
&&\overline{w}_{i}=\frac{\partial _{i}(\ r(\widetilde{\xi })\sin \theta (%
\widetilde{\xi },\vartheta )\sqrt{|q(\widetilde{\xi })|}\partial _{\varphi }[%
\overline{\chi }_{4}\varpi ])}{e^{\varpi }r(\widetilde{\xi })\sin \theta (%
\widetilde{\xi },\vartheta )\sqrt{|q(\widetilde{\xi })|}\partial _{\varphi
}\varpi }=\partial _{i}\overline{A}(\widetilde{\xi },\vartheta ,\varphi ),
\end{eqnarray*}%
for $x^{i}=(\widetilde{\xi },\vartheta ).$ With respect to N--adapted
frames, we model an ellipsoidal configuration with $\ r_{+}(\widetilde{\xi }=%
\widetilde{\xi }_{+})\simeq \frac{2\ \overline{M}(\ \widetilde{\xi }_{+})}{%
1+\varepsilon \underline{\zeta }\sin (\omega _{0}\varphi +\varphi _{0})}$,
for a corresponding value $\widetilde{\xi }_{+},$ constants $\underline{%
\zeta },\omega _{0}$ and $\varphi _{0}$ and eccentricity $\varepsilon .$

Putting together above formulas, we obtain{\small
\begin{eqnarray}
\mathbf{ds}^{2} &=&e^{\tilde{\psi}(\widetilde{\xi },\theta )}(d\widetilde{%
\xi }^{2}+\ d\vartheta ^{2})+  \label{ellipswh} \\
&&\frac{[\partial _{\varphi }\widetilde{\varpi }]^{2}}{\ \mu _{g}^{2}\
\lambda }(1+\varepsilon \frac{\partial _{\varphi }[\overline{\chi }_{4}%
\widetilde{\varpi }]}{\partial _{\varphi }\widetilde{\varpi }})\ ^{0}%
\overline{h}_{3}[d\varphi +\partial _{\widetilde{\xi }}(\ ^{\eta }\widetilde{%
A}+\varepsilon \overline{A})d\widetilde{\xi }+\partial _{\vartheta }(\
^{\eta }\widetilde{A}+\varepsilon \overline{A})d\vartheta ]^{2}  \notag \\
&&-\frac{e^{2\widetilde{\varpi }}}{4\mu _{g}^{2}|\ \lambda |}[1+\varepsilon
\overline{\chi }_{4}(\widetilde{\xi },\varphi )]e^{2B(\widetilde{\xi }%
)}[dt+\partial _{\widetilde{\xi }}(\ ^{\eta }n+\varepsilon \overline{n})~d%
\widetilde{\xi }+\partial _{\vartheta }(\ ^{\eta }n+\varepsilon \overline{n}%
)~d\vartheta ]^{2}.  \notag
\end{eqnarray}%
}If the generating functions $\widetilde{\varpi }$ and effective source $%
\lambda $ in massive gravity are such way chosen that polarization functions
(\ref{polfew}) can be approximated $\widetilde{\eta }_{a}\simeq 1$ and $%
^{\eta }\widetilde{A}$ and $\ ^{\eta }n$ are "almost constant", with respect
to certain systems of radial coordinates, the metric (\ref{ellipswh}) mimic
small rotoid wormhole like configurations with off--diagonal terms and $f$%
--modifications of the diagonal coefficients. It is possible to chose such
integration functions and constants that this class of stationary solutions
define wormhole like metrics depending generically on three space
coordinates with self--consistent "imbedding" in an effective massive
gravity background.

For more general classes of nonholonomic deformations, we can preserve
certain rotoid type symmetries but the "wormhole character" of solutions
became less clear.

\subsection{Solitonic waves for wormholes and black ellipsoids}

Let us consider two examples of gravitational solitonic deformations in
massive $f$--modified gravity.

\subsubsection{Sine--Gordon two dimensional nonlinear waves}

An interesting class of off--diagonal solutions depending on all spacetime
coordinates can be constructed by designing a configuration when a
1--solitonic wave preserves an ellipsoidal wormhole configuration. Such a
spacetime metric can be written in the form
\begin{eqnarray}
\mathbf{ds}^{2} &=&e^{\tilde{\psi}(x^{i})}(d\widetilde{\xi }^{2}+\
d\vartheta ^{2})+\omega ^{2}(\widetilde{\xi },t)\times ~  \label{solitwh} \\
&&\left[ \widetilde{\eta }_{3}(1+\varepsilon \frac{\partial _{\varphi }[%
\overline{\chi }_{4}\widetilde{\varpi }]}{\partial _{\varphi }\widetilde{%
\varpi }})\ ^{0}\overline{h}_{3}(\delta \varphi )^{2}-\widetilde{\eta }%
_{4}[1+\varepsilon \overline{\chi }_{4}(\widetilde{\xi },\varphi )]\ ^{0}%
\overline{h}_{4}(\delta t)^{2}\right] ,  \notag \\
\delta \varphi &=&d\varphi +\partial _{i}(\ ^{\eta }\widetilde{A}%
+\varepsilon \overline{A})dx^{i},\delta t=dt+~_{1}n_{i}(\widetilde{\xi }%
,\vartheta )dx^{i},  \notag
\end{eqnarray}%
for $x^{i}=(\widetilde{\xi },\vartheta )$ and $y^{a}=(\varphi ,t).$ The
"vertical" conformal factor
\begin{equation}
\omega (\widetilde{\xi },t)=4\arctan e^{m\gamma (\widetilde{\xi }-vt)+m_{0}},
\label{solit1}
\end{equation}%
where $\gamma ^{2}=(1-v^{2})^{-1}$ and constants $m,m_{0},v,$ defines a
1--soliton solution of the sine--Gordon equation  $\frac{\partial ^{2}\omega
}{\partial t^{2}}-\frac{\partial ^{2}\omega }{\partial \widetilde{\xi }^{2}}%
+\sin \omega =0$.

For $\omega =1,$ the metrics (\ref{solitwh}) are of type (\ref{ellipswh}). A
nontrivial value $\omega $ depends on the time like coordinate $t$ and has
to be constrained to conditions of type (\ref{confeq}), which can be written
for $~_{1}n_{2}=0$ and $~_{1}n_{1}=const$ in the form $\frac{\partial \omega
}{\partial \widetilde{\xi }}-~_{1}n_{1}\frac{\partial \omega }{\partial t}=0$%
. A gravitational solitonic wave (\ref{solit1}) will propagate
self--consistently in a rotoid wormhole background with $_{1}n_{1}=v$ which
solve both the sine--Gordon and constraint equations. Re--defining the
system of coordinates with $x^{1}=\widetilde{\xi }$ and $x^{2}=\theta ,$ we
can transform any nontrivial $~_{1}n_{i}(\widetilde{\xi },\theta )$ into
necessary $_{1}n_{1}=v$ and $_{1}n_{2}=0.$

\subsubsection{Three dimensional solitonic waves}

In general, we can construct various types of vacuum gravitational 2-d and
3-d configurations characterized by solitonic hierarchies and related
bi--Hamilton structures, for instance, of Kadomtsev--Petivashvili, KdP,
equations \cite{kadom} with possible mixtures with solutions for 2-d and 3-d
sine--Gordon equations etc, see details in Ref. \cite{vacarsolitonhier}.

Let us consider a solution of KdP equation for the $v$--conformal factor $%
\omega =\check{\omega}(\widetilde{\xi },\varphi ,t),$ when $y^{4}=t$ is
taken as a time like coordinate, as
\begin{equation}
\pm \check{\omega}^{\ast \ast }+(\partial _{t}\check{\omega}+\check{\omega}\
\check{\omega}^{\bullet }+\epsilon \check{\omega}^{\bullet \bullet \bullet
})^{\bullet }=0,  \label{kdp1}
\end{equation}%
with dispersion $\epsilon .$ In the dispersionless limit $\epsilon
\rightarrow 0,$ we can consider that the solutions are independent on $%
\varphi $ and determined by Burgers' equation $\partial _{t}\check{\omega}+%
\check{\omega}\ \check{\omega}^{\bullet }=0.$ For 3--d solitonic
configurations, the conditions (\ref{confeq}) are written in the form  $%
\mathbf{e}_{1}\check{\omega}=\check{\omega}^{\bullet }+w_{1}(\widetilde{\xi }%
,\vartheta ,\varphi )\check{\omega}^{\ast }+n_{1}(\widetilde{\xi },\vartheta
)\partial _{t}\check{\omega}=0$.  If $\check{\omega}^{\prime }=0,$ we can
fix $w_{2}=0$ and $n_{2}=0.$

Such solitonic deformations of the wormhole metrics and their massive
gravity and $f$--modifications can be parameterized in the form
\begin{eqnarray*}
\ _{{}}\mathbf{g} &=&e^{\psi (\widetilde{\xi },\vartheta )}(d\widetilde{\xi }%
\otimes d\widetilde{\xi }+d\vartheta \otimes d\vartheta )+\left[ \check{%
\omega}(\widetilde{\xi },\varphi ,t)\right] ^{2}h_{a}(\widetilde{\xi }%
,\varphi )\ \mathbf{e}^{a}\otimes \mathbf{e}^{a}, \\
\mathbf{e}^{3} &=&d\varphi +w_{1}(\widetilde{\xi },\vartheta ,\varphi )d%
\widetilde{\xi },\ \mathbf{e}^{4}=dt+n_{1}(\widetilde{\xi },\vartheta )d%
\widetilde{\xi }.
\end{eqnarray*}%
This class of metrics does not have (in general) Killing \ symmetries but
may possess symmetries determined by solitonic solutions of (\ref{kdp1}).

In a similar form, we can construct solutions for any $\check{\omega}$
defined by any 3--d solitonic and/ or other nonlinear wave equations, or
generate solitonic deformations for $\omega =\check{\omega}(\vartheta
,\varphi ,t).$

\subsection{Ringed wormholes}

Using the AFDM, we can generate ansatz for a rotoid wormhole plus a torus
(ring) configuration, {\small
\begin{eqnarray}
\mathbf{ds}^{2} &=&e^{\tilde{\psi}(x^{i})}(d\widetilde{\xi }^{2}+\
d\vartheta ^{2})+\widetilde{\eta }_{3}(1+\varepsilon \frac{\partial
_{\varphi }[\overline{\chi }_{4}\widetilde{\varpi }]}{\partial _{\varphi }%
\widetilde{\varpi }})\ ^{0}\overline{h}_{3}(\delta \varphi )^{2}  \notag \\
&& -F(\widetilde{\xi },\vartheta ,\varphi )\widetilde{\eta }%
_{4}[1+\varepsilon \overline{\chi }_{4}(\widetilde{\xi },\varphi )]\ ^{0}%
\overline{h}_{4}(\delta t)^{2}  \notag \\
\delta \varphi &=&d\varphi +\partial _{i}(\ ^{\eta }\widetilde{A}%
+\varepsilon \overline{A})dx^{i},\delta t=dt+~_{1}n_{i}(\widetilde{\xi }%
,\vartheta )dx^{i},  \label{torusm}
\end{eqnarray}%
} for $x^{i}=(\widetilde{\xi },\vartheta )$ and $y^{a}=(\varphi ,t),$ where
the function $F(\widetilde{\xi },\vartheta ,\varphi )$ in conventional
spherical coordinates can be rewritten equivalently in conventional
Cartesian coordinates as  $F(\widetilde{x},\widetilde{y},\widetilde{z}%
)=\left( R_{0}-\sqrt{\widetilde{x}^{2}+\widetilde{y}^{2}}\right) ^{2}+%
\widetilde{z}^{2}-a_{0}$, for $a_{0}<a,R_{0}<r_{0}$. We get a ring around
the wormhole throat \footnote{%
we can consider wormholes in the limit $\varepsilon \rightarrow 0$ and for
corresponding approximations $\widetilde{\eta }_{a}\simeq 1$ and $^{\eta }%
\widetilde{A}$ and $\ ^{\eta }n$ to be almost constant}. The ring
configuration is defined by the condition $\ F=0.$ For $F=1,$ we get a
metric of type (\ref{solitwh}) with $\omega =1.$ In above formulas, $R_{0}$
is the distance from the center of the tube to the center of the torus/ring
and $a_{0}$ is the radius of the tube.

If the wormhole objects exist, the variants ringed by a torus may be stable
for certain nonholonomic geometry and exotic matter configurations. We omit
in this work a rigorous stability analysis as well a study of issues related
to cosmic censorship criteria etc.

\subsection{Modified wormholes with induced torsion}

The examples for wormhole nonholonomic deformations considered above are for
effective LC--configurations which can be effectively modelled by nonlinear
off--diagonal interactions in GR. Here, we provide an example of a class of
stationary off--diagonal solutions with nontrivial torsion effects resulting
in additional effective rotation proportional to $\mu _{g},$ see a similar
configuration (\ref{ofindtmg}). \ The corresponding off--diagonal quadratic
element is given by
\begin{eqnarray}
ds^{2} &=&e^{\tilde{\psi}(\widetilde{\xi },\vartheta )}(d\widetilde{\xi }%
^{2}+\ d\vartheta ^{2})+\frac{(\partial _{\varphi }\Phi )^{2}}{\ \mu
_{g}^{2}\lambda (\widetilde{\xi },\vartheta )\Phi ^{2}}[dy^{3}+\frac{%
\partial _{i}\Phi }{\partial _{\varphi }\Phi }dx^{i}]^{2}  \label{whts} \\
&&-\frac{\Phi ^{2}}{4\mu _{g}^{2}|\lambda (\widetilde{\xi },\vartheta )|}%
[dt+(\ _{1}n_{k}(\widetilde{\xi },\vartheta )+\ _{2}n_{k}(\widetilde{\xi }%
,\vartheta )\frac{4\mu _{g}(\partial _{\varphi }\Phi )^{2}}{\Phi ^{5}}%
)dx^{k}]^{2},  \notag
\end{eqnarray}%
for $x^{i}=(\widetilde{\xi },\vartheta )$ and generating function $\Phi
=\exp [2\widetilde{\varpi }(\widetilde{\xi },\vartheta ,\varphi )].$ The
d--torsion coefficients (\ref{dtors}) for this metric are not trivial if $\
_{2}n_{k}\neq 0.$ This and other setting for more general sources $\Upsilon (%
\widetilde{\xi },\vartheta ,\varphi )$ (\ref{source}) different classes of
N--coefficients lead to different characteristic geometric and physical
properties which are very different from LC--configurations.

We can parameterize (\ref{whts}) in any form (\ref{masds}), (\ref{offdwans1}%
), (\ref{ellipswh}), (\ref{solitwh}) in order to generate off--diagonal
solutions with $\mu _{g}$-- and/of $f$--modifications possessing rotoid
and/or solitonic symmetries characterized by nonholonomic torsion. If a
vertical conformal factor $\omega $ similar to (\ref{solitwh}) is
considered, the metric and induced torsion fields might depend on all four
spacetime coordinates. Toroidal configurations of type (\ref{torusm}) can be
constructed if a toroidal function of type $F(\widetilde{x},\widetilde{y},%
\widetilde{z})$ is introduced before the $v$--components of metrics in (\ref%
{whts}).

\section{Concluding Remarks and Discussion}

\label{s6} Modified gravity theories with functional dependence on curvature
and other traces of energy--momentum tensors for matter fields, torsion
sources etc and/or with contributions by massive/bi--metric and generalized
connection terms for Lagrangians belong to the most active research area
oriented to solution of important problems in modern cosmology and particle
physics. As we can see in recent literature, many interesting and original
classical and quantum scenarios can be elaborated for naive additions of
mass terms, nontrivial geometric backgrounds and nonlinear interactions via
polarized constants and quantum corrections. Such constructions are grounded
on geometric models and solutions for certain (generalized) effective
Einstein equations with high degrees of symmetries (for Killing vectors) and
diagonalizable metrics.

In our research, we are focused on exact and approximate generic
off--diagonal solutions in gravity theories with generalized symmetries and
dependencies via generating and integration functions on much as possible
spacetime coordinates (for instance, on three and four ones on 4--d
manifolds). It is a difficult mathematical task to construct such solutions
in analytic form and to provide and study certain physical important
examples and interesting effects relates to outstanding issues, for
instance, in cosmology and astrophysics. Nevertheless, all candidates to
gravity theories are characterised by complex off--diagonal systems of
nonlinear partial equations and fundamental classical and quantum properties
of gravitational and matter fields interactions should be studied with
regard to the found most  general classes of solutions and nonlinear
nonholonomically constrained configurations. Here we note that although the
equations of modified gravity theories are rather involved, they became much
simple in certain adapted system of reference and certain types of
nonholonomic constraints on certain classes for generic off--diagonal
solutions. The surprising thing is that in many cases, for well--defined
geometric conditions, we can model certain classes of nonlinear solutions
both in an effective Einstein like theory (with off--diagonal metrics and
generalized, or the Levi--Civita, LC, connection) and in modified
bi--metric/--connection gravity models, in general, with nontrivial mass
terms. Hence, a generic off--diagonal solution with arbitrary generating and
integration functions and constants in GR can be regarded as a possible
analog of various types of similar solutions in modified gravity theories.
In many cases, we can argument a quite conservative opinion: may be it is
not necessary to modify the "canonical" Einstein gravity if in the framework
of this theory we are able to explain many fundamental issues and observable
cosmological effects via certain generalized off--diagonal solutions with
generic off--diagonal interactions and nonholonomic constraints?

In order to investigate certain physical implications of off--diagonal
solutions and the possibility to mimic physical effects in one theory by
effective analogs of such solutions in another class of theories, a general
geometric/analytic method of constructing exact solutions should be applied.
For such purposes, we developed the anholonomic frame deformation method,
AFDM, see \cite{vadm1,vadm2,vw1,vw2,vw3} and references therein. Following
such a geometric method, various classes of gravitational and matter field
equations in \ modified gravity (MG) and Einstein gravity theories can be
decoupled and integrated in very general forms if necessary types of adapted
frame and connection structures are considered. We can impose constraints,
at the end, for extracting Levi Civita configurations. This way, a wide
variety of generalized locally anisotropic wormhole and matched exterior
black holes can be constructed. They can be derived for certain exotic
matter and/or for off--diagonal configurations of metrics describing
nonlinear gravitational and matter field interactions which may limit
certain de Sitter spacetimes with effective "anisotropically" polarized
cosmological and matter fields constants. The assumption on stationary
properties of such locally anisotropic spacetimes is introduced from the
very beginning even solitonic waves can be involved. Note that the method
allows us to find a wide variety of non--stationary exact solutions.

In this paper, we focused on generic off--diagonal solutions which are
constructed as nonholonomic deformations of pseudo-Riemannian metrics with
two Killing vectors (in particular, they can be solutions of the Einstein
equations) into certain classes of generalized exact solutions in massive
gravity with possible small parametric deformations related to $f$--modified
gravity. We proved that one exists a formal integration procedure via
effective polarization functions which allows us to construct various
classes of exact solutions depending generically on three and four
coordinates on (generalized) four dimensional spacetimes. In explicit form,
we constructed and studied off--diagonal deformations of wormhole solutions
matching exterior (in general, nonholonomically deformed) de Sitter
spacetimes with contributions by nontrivial massive gravitational terms and
ellipsoidal $f$--modifications of de Sitter metrics. We also analysed
solitons waves, possible "ringed wormhole" like configurations, modified
wormholes "distorted" in nonholonomically induced torsion etc.

There is still much to be learned about possibilities of the AFDM and
possible relations of such way constructed off--diagonal solutions with
massive gravity, $f$--modified, Finsler like etc theories. Here, it should
be noted that such nonholonomic structures were originally considered in
Finsler like theories, fractional generalizations etc to modern cosmological
scenarios \cite{stavr,mavr,castro,calcagni}. This paper and discussion
provided just a glimpse to potential applications and future our work.

\vskip5pt

\textbf{Acknowledgments:\ } The work is partially supported by the Program
IDEI, PN-II-ID-PCE-2011-3-0256. The author is grateful to organizers and
participants of respective sections at ERE 2012 and MG13, where some results
of this paper where presented. He also thanks S. Capozziello,  P. Stavrinos, M. Visinescu, D. Singleton and N.
Mavromatos  for important discussions, critical
remarks and substantial support.

\end{document}